\documentclass[stuctureabstract]{aa}  
 \usepackage{tabularx}                                  % (traditional abstract) 
\usepackage{graphicx}
\usepackage{amsmath}
\usepackage{txfonts}
\usepackage{booktabs}
\newcommand{\ra}[1]{\renewcommand{\arraystretch}{#1}}
\usepackage[sort]{natbib}
\begin{document}
   \title{Tully-Fisher analysis of the multiple cluster system Abell 901/902\thanks{Based on observations with the European Southern Observatory Very Large Telescope (ESO-VLT), observing run ID 384.A-0813.}}

%   \subtitle{Spectroscopic redshifts, velocity dispersions, rotation curve asymmetries, morphological descriptors}

   \author{Benjamin B\"{o}sch\inst{1}\fnmsep\thanks{\email{benjamin.boesch@uibk.ac.at}}
          \and Asmus B\"{o}hm\inst{1}
          \and Christian Wolf\inst{2}
          \and Alfonso Arag\'{o}n-Salamanca\inst{3}
          \and Bodo L. Ziegler\inst{4}
          \and Marco Barden\inst{1}
          \and Meghan E. Gray\inst{3}
          \and Michael Balogh\inst{5}
          \and Klaus Meisenheimer\inst{6}
          \and Sabine Schindler\inst{1}
          }

   \institute{
Institute for Astro- and Particle Physics, University of Innsbruck, Technikerstr. 25/8, A-6020 Innsbruck, Austria
\and Department of Physics, Denys Wilkinson Building, University of Oxford, Keble Road, Oxford OX1 3RH, UK
 \and School of Physics and Astronomy, The University of Nottingham, University Park, Nottingham NG7 2RD
 \and Department of Astronomy, University of Vienna, T\"{u}rkenschanzstr. 17, 1180 Wien, Austria       
 \and Department of Physics, University of Waterloo, Waterloo, ON N2L 3G1, Canada
 \and Max Planck Institute for Astronomy, K\"{o}nigstuhl 17, 69117 Heidelberg, Germany
  }          
   \date{Received 25 March 2013 / Accepted 19 April 2013}

% \abstract{}{}{}{}{} 
% 5 {} token are mandatory
 
  \abstract
  % leave it empty if necessary  
   {%This work is incorporated into the STAGES (=\textbf{S}pace \textbf{T}elescope \textbf{A}bell 901/902 \textbf{G}alaxy \textbf{E}volution \textbf{S}urvey) collaboration, which is designed to probe the physical drivers of galaxy evolution across the A901/902 supercluster.
   }
   {We derive rotation curves from optical emission lines of 182 disk galaxies (96 in the cluster and 86 in the field) in the region of Abell 901/902 located at $z\sim 0.165$. We continue the kinematic analysis presented in a previous paper. Here, we focus on the analysis of B-band and stellar-mass Tully-Fisher relations. We examine possible environmental dependencies and differences between normal spirals and "dusty red" galaxies, i.e. disk galaxies that have red colors due to relatively low star formation rates.
   }
   {Assuming a tilted ring model we simulate the spectroscopy of a given galaxy and reconstruct the maximum rotation velocity from the best-fitting simulated rotation curve. We fit regression lines adopting a maximum likelihood method based on Bayes' theorem to build B-band and stellar-mass Tully-Fisher relations. 
   }
   {We find no significant differences between the best-fit TF slope of cluster and field galaxies. At fixed slope, the field population with high-quality rotation curves (57 objects) is brighter by $\Delta M_{B}=-0\fm42\pm0\fm15$ than the cluster population (55 objects). We show that this slight difference is at least in part an environmental effect. The scatter of the cluster TFR increases for galaxies closer to the core region, also indicating an environmental effect. Interestingly, dusty red galaxies become fainter towards the core at given rotation velocity (i.e. total mass). This indicates that the star formation in these galaxies is in the process of being quenched. The luminosities of normal spiral galaxies are slightly higher at fixed rotation velocity for smaller cluster-centric radii. Probably these galaxies are gas-rich (compared to the dusty red population) and the onset of ram-pressure stripping increases their star-formation rates.\\
Galaxies of smooth morphology show higher rms values in the fitting of their rotation curves. This is particularly the case for dusty red galaxies. A cluster-specific interaction process like ram-pressure stripping is the best explanation, since it only affects the gaseous disk but not the stellar morphology.
}
   {The results from the TF analysis are consistent with and complement our previous findings. Dusty red galaxies might be an intermediate stage in the transformation of infalling field spiral galaxies into cluster S0s, and this might explain the well-known increase of the S0 fraction in galaxy clusters with cosmic time.}

   \keywords{galaxies: clusters: general --
                galaxies: clusters: individual (A901, A902) --
                galaxies: evolution --
                galaxies: kinematics
               }
               
\authorrunning{B. B\"{o}sch et al.} 
\titlerunning{Tully-Fisher analysis of A901/902}

   \maketitle
%
%________________________________________________________________

\section{Introduction}
The empirically found, tight correlation between the rotation velocity $V_{\mathrm{rot}}$ of disk galaxies and their luminosity $L$ ($L \propto V^{\alpha}_{\mathrm{rot}}$) was originally introduced by \citet{tully77} as a new method of determining distances to galaxies. Since the maximum rotation velocity is proportional to the total mass of a galaxy and the luminosity correlates with stellar mass, the Tully-Fisher relation (TFR) provides a link between the dark and visible matter component of a galaxy. This link is assumed to be a consequence of the formation scenario of disk galaxies within the framework of a cold dark matter universe \citep{blumenthal86}. Initially in equilibrium with the dark matter halo, rotating disks form from collapsing gas clouds. In the process, angular momentum is conserved. In contrast to dissipationless dark matter, baryonic particles can loose part of their energy via radiation and hence are able to form more compact structures \citep[e.g.][]{dutton07}.\\
Analysing the Tully-Fisher relation in different luminosity bands \citep[e.g.][]{kannappan02,pizagno07,verheijen01} or plotting the rotation velocity directly against the stellar mass \citep[smTFR; e.g.][]{pizagno05,kassin07,miller11} or baryonic mass \citep[bTFR; e.g.][]{verheijen01,puech10,mcgaugh00} can give valuable insight into the formation history of spiral galaxies. Comparing the best-fit parameters of the TFR (slope, intercept and scatter) at different redshifts to predictions of numerical simulations helps to constrain the physical processes important for the evolution of spiral galaxies \citep[e.g.][]{vogt96,boehm04,boehm07,bamford05,bamford06,kassin07,puech08,miller11,miller12}. \\
The well-known morphology-density relation \citep{dressler80} has been confirmed in many subsequent studies highlighting a picture of high-density environments altering the properties of galaxies. Besides a prevalence of more early-type morphology, galaxies residing in clusters on average have e.g. redder colours \citep[e.g.][]{blanton05}, lower star-formation rates \citep[e.g.][]{verdugo08,lewis02} and a reduced gas content \citep[e.g.][]{giovanelli85,solanes01} compared to field galaxies. Consequently the Tully-Fisher parameters might vary between different environments and hence are the topic of several studies \citep[e.g.][]{mj03,nakamura06,bamford05,moran07,ziegler03,kutdemir08,kutdemir10,jaeger04}. \\
Some of them show conflicting results and conclusions. The main reason for this are differences in the data analysis including the observation of the target galaxies, the derivation of luminosities and rotation velocities, or the methods used to fit the TFR. Several definitions of rotation velocities exist and many groups use slightly different methods to account for observational effects and to infer the intrinsic rotation curve of a galaxy. Different sample sizes, selection criteria and galaxy populations make a  comparison of results even more complicated.\\
However, several general trends could be established. The slope of the TFR increases towards redder photometric passbands \citep[e.g.][]{kannappan02,pizagno07,verheijen01}. Redder bands are less sensitive to recent star-formation events and dust extinction and better trace the underlying old stellar population, which dominates the stellar mass. On the one hand, the B-band TFR shows a brightening of the intercept up to $\Delta M_{B} \sim 1$ mag out to a redshift of $z\sim1$ \citep[e.g.][]{vogt96,boehm04,weiner06,miller11,ziegler02,fernando10}. This is most likely due to an enhancement of the star-formation rate and a younger stellar population \citep{hopkins04}. On the other hand, there is little or no evolution in the stellar-mass or baryonic Tully-Fisher relation \citep[e.g.][]{flores06,miller11,puech08}. In agreement with simulations \citep[e.g.][]{steinmetz99}, most studies measure a larger scatter for TFRs at higher redshifts \citep[e.g.][]{bamford06,boehm07,kassin07}.\\
 Comparisons between the TFRs of cluster and field galaxies show no clear results. While \citet{jaffe11}, \citet{nakamura06} and \citet{ziegler03} find no significant differences between these two environments, \citet[][]{bamford05} and \citet[][]{mj03} deduce higher B-band luminosities in clusters compared to the field. \citet{moran07} find a larger scatter for cluster galaxies. Again, sample selection and analysis techniques might influence the results. But also the choice of a particular cluster and its ICM properties and dynamical state might play a role.\\
Using spatially resolved spectra obtained with VIMOS (VIsible MultiObject Spectrograph) on the VLT (Very Large Telescope) in combination with already existing data from the STAGES (Space Telescope A901/902 Galaxy Evolution Survey) project \citep{gray09}, we continue the kinematic analysis of \citet[][hereafter Paper I]{boesch13}. We exploit rotation curves of disk galaxies in and around the multiple cluster system Abell 901/902 with a focus on the analysis of the Tully-Fisher relation. The paper is organised as follows. Sect. 2 gives a short introduction into the STAGES project, outlines the sample selection and observations and reviews results of Paper I. Sect. 3 describes the derivation of the parameters needed for the Tully-Fisher analysis and presents adopted methods to build the TFR. Sect. 4 contains the TF analysis, which provides the basis for the residual analysis of Sect. 5. Finally, Sect. 6 summarises the main points.\\
  Throughout this paper, we assume a cosmology with $\Omega_{m}=0.3$, $\Omega_{\Lambda}=0.7$ and $H_{0}=70$ $\mbox{kms}^{-1}\mbox{Mpc}^{-1}$. Magnitudes quoted in this paper are in the AB system. 

%__________________________________________________________________

\section{Data}
\subsection{The STAGES data set}
STAGES is a multi-wavelength survey centred on the area around the multiple cluster system Abell 901/902. It comprises four subclusters: A901a (at redshift 0.164), A901b (z=0.163), A902 (z=0.167) and the SW group (z=0.169). They are in a non-virialised state \citep{boesch13} most likely coalescing to one big system in the future. The $0\fdg5\times0\fdg5$ ($\sim5\times5 \textrm{Mpc}^{2}$) region of Abell 901/902 has been the subject of Hubble Space Telescope/Advanced Camera for Surveys (HST/ACS) observations producing an 80-tile mosaic of V-band (F606W) images. We have at hand 17-band COMBO-17 (Classifying Objects by Medium-Band Observations) ground-based imaging \citep{wolf03}, Spitzer 24 $\mu m$, and XMM-Newton X-ray data \citep{gilmour07}. Gravitational lensing maps \citep{heymans08}, spectral energy distributions \citep{wolf03}, stellar masses \citep{borch06}, star-formation rates \citep{bell07}, mean stellar ages and morphological classifications provide an excellent starting point for further analysis. \citet{gray09} give an overview of the STAGES project and present the publicly available master catalogue \citep{stagescat09}.
%__________________________________________________________________
\subsection{Observation, sample selection \& data reduction}
Information on the spectroscopic observations of A901/902 with VLT/VIMOS as well as details on the data reduction process are presented in Paper I. Thus, we will only give a short review of the most important steps here.\\
Multi-object spectroscopy (MOS) with VIMOS was carried out between February 8 and March 10, 2010 (ESO-ID 384.A-0813, P.I. A. B\"{o}hm). Four VIMOS pointings ($\sim 1'$ overlap) sample the whole A901/902 field (two exposures of 1800 seconds each). We opted for the high resolution grism \textit{HR-blue} with a slit width of $1\farcs2$, which covered a $\sim$2050$\AA$ spectral range in the wavelength interval $[3700\AA,6740\AA]$. This configuration yielded a spectral resolution of $R\sim 2000$ and an average dispersion of $0.51\AA /$pixel at an image scale of $0\farcs205/$pixel. The robustness of our analysis against varying seeing conditions during spectroscopy are discussed in Sect. \ref{sec:rcm}.\\
Our primary targets for spectroscopy are disk galaxies showing emission lines. Hence, we selected galaxies with stellar masses $M_{*}>10^{9} M_{\odot}$, absolute magnitudes $M_{B}<-18$, a star-forming SED (selection criterion in the public catalogue \citet{stagescat09}: $\mathrm{sed\_type} \ge 2$) and a visually confirmed disk component on the ACS images with an inclination angle $i>30^{\circ}$. We require photometric redshifts in the interval $0.1<z_{\mathrm{phot}}<0.26$ resulting in a matched cluster and field sample. This selection resulted in $\sim$320 cluster and $\sim$160 field candidates. Using the ESO software VMMPS (VIMOS Mask Preparation Software), we manually placed MOS slits along the apparent major axis of the galaxies (allowing slit tilt angles up to $\pm 45^{\circ}$). In the end, we obtained spectra of $215$ different galaxies.\\
One science driver of our project is to investigate the population of \textit{dusty red galaxies} \citep{wolf05}. In total, $44$ of the $215$ spectra belong to galaxies with this SED type (selection criterion in the public catalogue \citet{stagescat09}: $\mathrm{sed\_type} = 2$). Dusty red galaxies are spread across the red sequence in the U-V colour-magnitude diagram. But, unlike early-type red galaxies, part of their colour is due to intrinsic dust extinction ($\mathrm{E(B-V)}\geq 0.1$). Hence, they could be interpreted as the low specific-SFR tail of the blue cloud \citep{wolf05}.\\
We performed the spectroscopic data reduction mainly using the ESO-VIMOS pipeline (version 2.2.3). The main reduction steps were bias subtraction, correction of distortions resulting from the focal reducer using flat field exposures (a potential normalisation was skipped due to reflections present in the flat-field images) and wavelength calibration. Additionally, we performed bad pixel and cosmic ray cleaning. Wavelength calibration for spectra with slits in the upper third of the CCD or with large tilt angles was improved by adding Argon lines to the catalogue of arc lamp lines. For more details see Paper I.\\

\subsection{Results from Paper I}
In this subsection we will review results from \citet{boesch13}, which will be important for the current analysis and help to interpret our data.\\
We were able to measure spectroscopic redshifts for 200 different galaxies, either from emission lines (188) or from strong absorption lines (12). Using galaxies observed more than once and following a prescription outlined in \citet{mj08}, we estimated the typical redshift error to be $\delta_{z}/(1+z) \sim 0.00016 $ or  $47 \mathrm{km/s}$ in rest-frame.\\
 \begin{figure}[]
   \centering
   \includegraphics[angle=0,width=\columnwidth]{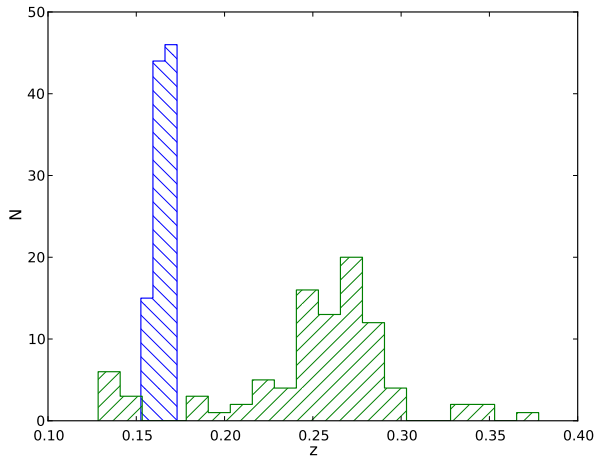}
      \caption{Redshift histogram of the 200 galaxies of our sample. The cluster sample (106 objects; blue, ``backslash" hatching) peaks at the cluster redshift $z\sim0.165$. The field sample (94 objects; green, ``slash" hatching) shows a “bump” at $z\sim0.26$, which might be a volume effect. 
              }
         \label{fig:zhist}
   \end{figure}  
We used the biweight estimator of scale and location \citep{beers90} to estimate the velocity dispersion and mean redshift of each of the four subclusters individually. On the basis of these properties, we determined cluster membership using a 3-$\sigma$-clipping. We identified 106 cluster members. Fig. \ref{fig:zhist} shows the redshift histogram of the whole sample, separated into cluster and field galaxies. \\
 We found that the fraction of cluster galaxies with distorted rotation curves is $75\%$ higher than for field galaxies. Most of these differences can be attributed to morphologically undistorted galaxies with a relatively smooth stellar disk. Dusty red galaxies account for a large part of this population. This suggests a cluster-specific mechanism, which mainly affects the gas disk of the cluster spirals, with ram-pressure stripping as the best candidate. Investigations of the gas concentration implied that most spiral galaxies traverse the cluster for the first time. Especially for dusty red galaxies, ram-pressure could be a dominant mechanism in A901/902, even in the cluster outskirts.  

\section{Methods}
Most galaxy properties such as stellar masses, apparent magnitudes and inclination angles are already provided by the STAGES project \citep{stagescat09}. Assuming a \citet{kroupa93} stellar initial mass function, \citet{borch06} derived stellar masses from an SED-fitting of templates from PEGASE \citep{fioc97} population synthesis models to the COMBO-17 fluxes. They noted that alternatively using \citet{kroupa01} or \citet{chabrier03} initial mass functions would result in the same stellar masses to within $\sim 10 \%$.\\
We only need to determine rest-frame absolute magnitudes and maximum rotation velocities for each object, and then derive the Tully-Fisher relations for several samples. We adopt $V_{2.2}$, the velocity at 2.2 disk scalelengths $r_d$, as our kinematic measure, because it produces the tightest TFRs and is more robust against observational errors than other velocity definitions \citep{courteau97}. In contrast, the asymptotic velocity $V_{\mathrm{max}}$, e.g., is easily overestimated in galaxies with insufficiently extended rotation curves. We note, that the chosen point at $2.2r_d$ is the radius where the rotation velocity of an idealised self-gravitating disk is expected to peak \citep{freeman70}.
\subsection{Rest-frame magnitudes}
Apparent V-band magnitudes \textit{Mag$\_{best}$} from the HST/ACS (F606W) images were measured using SExtractor \citep{bertin96}. To compute absolute magnitudes in the B-band, we apply a k-correction $k_{B}(V,T,z)$, the distance modulus $D(z,H_{0},\Omega_{m},\Omega_{\Lambda})$ and a correction for Galactic absorption $A^{g}_{V}$ and intrinsic extinction $A^{i}_{B}(a/b,V_{\mathrm{max}})$.\\
The k-correction is needed to transform the apparent magnitude in an observed band to a rest-frame band for a given galaxy. This correction depends on the SED type T and the redshift $z$. In our case it is sufficient to use the transformation $V\rightarrow B$, since at the redshifts of our sample the V-band matches the rest-frame B-band. We determine the correct SED T-type individually by comparing our galaxy spectra to the templates in the Kennicutt catalogue \citep{kennicutt92a,kennicutt92b}. Besides the stellar continuum shape of the spectra, we use e.g. equivalent widths, the relative strength of emission lines, and the prominence of absorption lines for classification. The SED type T ranges from 1 (for a typical spectrum of an Sa galaxy) to 10 (for type Im). For more details see \citet{boehm04}.\\
%The distance modulus depends on the redshift $z$ of the galaxy as well as on the used cosmology. We assume a flat $\lambda$CDM Universe with $\Omega_{m}=0.3$, $\Omega_{\Lambda}=0.7$ and $H_{0}=70\mbox{kms}^{-1}\mbox{Mpc}^{-1}$.\\
 We adopt the correction factors for Galactic extinction in the direction of A901/902 ($\alpha$=$9\mathrm{h}56\mathrm{min}17\mathrm{s}$, $\delta$=$-10^{\circ}01'11''$, J2000) of the observed V-band from the NASA extragalactic database following \citet{schlafly11}. This yields the following value for the whole sample: $A^{g}_{V}=0\fm136$.\\
A critical step is the correction for intrinsic absorption $A^{i}_{B}(a/b,V_{\mathrm{max}})$ due to dust extinction. The intrinsic absorption depends on the inclination angle $i$ or equivalently the axis ratio $a/b$ of a galaxy and is maximal for an edge-on and minimal for a face-on configuration. We follow the prescription of \citet{tully98} and apply a correction, which is also correlated with the maximum rotation velocity $V_{\mathrm{max}}$ of a galaxy, since extinction is stronger in more massive galaxies \citep{giovanelli95}. This correction is relative to a face-on orientation and does not account for a potential residual absorption in a face-on system. All in all, we apply the following intrinsic absorption correction:
\begin{equation}
\begin{aligned}
A^{i}_{B}(a/b,V_{\mathrm{max}})&= \log(a/b)\left[-4.48+2.75\log(V_{\mathrm{max}})\right]\\
\end{aligned}
\label{eq:ia}
\end{equation}
To summarise, the absolute magnitude in a galaxy's rest-frame B-band is calculated as follows:
\begin{equation}
M_{B} = m_{V} - D(z) - k_{B}(V,T,z) - A^{g}_{V} - A^{i}_{B}(a/b,V_{\mathrm{max}})
\label{eq:mabs}
\end{equation}
We compute the corresponding errors of the absolute magnitudes via error propagation. We assume an uncertainty of $10\%$ for the axis ratio and an SED classification error of $\delta T = 2$. The latter transforms into an average uncertainty in the k-correction of $\sigma_{k}=0\fm08$.
 
\subsection{Rotation-curve extraction}
\label{sec:rce}
We address the extraction of the rotation curves from the emission lines of a spectrum only very briefly here, since this process is already described in detail in Paper I. Depending on a galaxy's redshift and position on the CCD, up to four emission lines are visible in the VIMOS spectra: $[O III] \,\lambda 5007\mathring{A}$, $[O III] \,\lambda 4959\mathring{A}$, $H\beta\,\lambda4861\mathring{A}$ and $[OII] \,\lambda 3726/3729$. Prior to the emission line fitting we add up three (or five, in cases with a very low S/N) neighbouring rows for each spatial position of the spectrum to enhance the S/N. We then transform red- and blueshifts of the emission lines into an observed rotation curve. The kinematic centre of a galaxy by definition has no Doppler shift. Its exact position significantly influences the shape of the rotation curve and must therefore be derived with care. Primarily, we determine the kinematic centre of a galaxy to within $0\farcs1$ by fitting a Gaussian to its (spatial) luminosity profile at the spectral position of the emission line, i.e. the luminous centre is assumed to be also the kinematic centre. In a second step, we slightly shift the centre by up to $\pm1.5$ pixels, minimising the rotation curve asymmetry adopting an asymmetry measure defined in \citet{dale01}). We extracted rotation curves from at least one emission line of 182 different galaxies (86 in the field and 96 in the cluster).
   
\subsection{Rotation-curve modelling}
\label{sec:rcm}
\begin{figure*}[]
   \centering
   \includegraphics[angle=0,width=\textwidth]{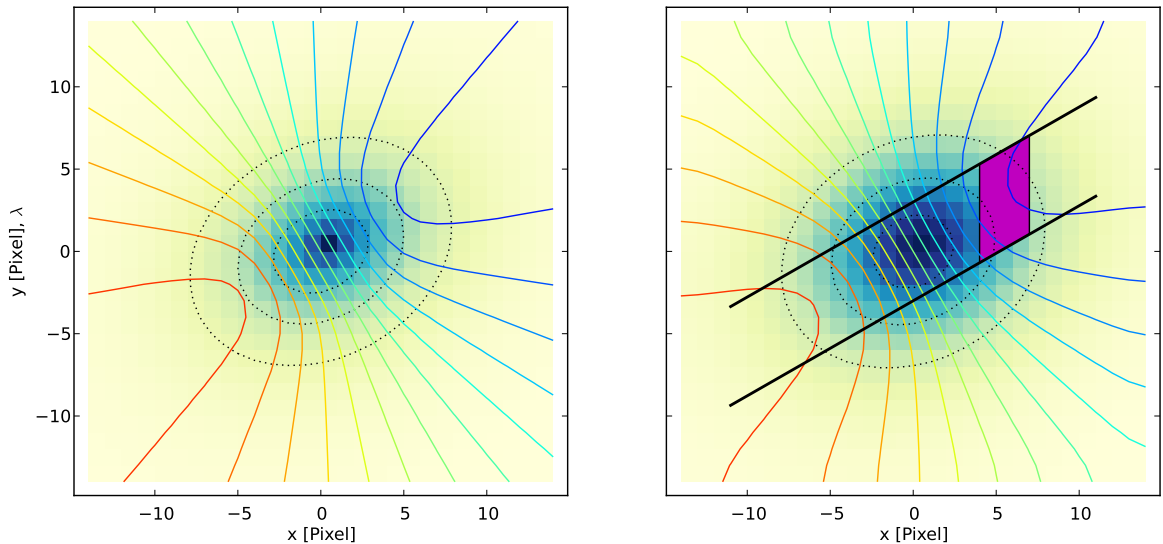}
      \caption{Simulated 2D intensity and velocity field of a galaxy. The fields are projected according to the inclination and position angle (here $i=40^{\circ}$, $\theta=30^{\circ}$). The direction of dispersion is along the y-axis. The exponential intensity profile is depicted in blue colour scale (here the scale length is $r_{d}=5.3$ pix). Isophotes corresponding to [50, 30, 15]\% of the central intensity are superimposed (black dotted ellipses). The coloured curves (``spider-diagram") indicate iso-velocity lines ranging from $-60\%V_{\mathrm{max}}$(red) to $60\%V_{\mathrm{max}}$(blue) in steps of $10\%$. Right panel: The intensity and velocity field is additionally convolved with a Gaussian Point Spread Function (FWHM = $0\farcs7=3.5$ pix). The slit (in black) is placed along the apparent major axis of the galaxy. At each spatial position along the x-axis, an intensity-weighted average of the velocity values inside the slit is calculated. The shaded magenta area indicates such a velocity range for a certain spatial position (with boxcar smoothing of $3$ pixels). Here, the kinematic and the luminous centre coincide.
              }
         \label{fig:sim1}
   \end{figure*}

At face value, an extracted rotation curve (RC) only traces the line-of-site rotation velocity of a galaxy. This observed position-velocity information does not represent the \textit{intrinsic} rotation of a disk galaxy. Several geometric and observational effects have to be taken into account. The inclination of the disk with respect to the line-of-sight, the position angle of the galaxy's apparent major axis with respect to the direction of dispersion, beam-smearing due to the integration along the slit width and seeing effects may ``disguise" the intrinsic shape of the rotation curve. For a detailed discussion see \citet{boehm04}.\\
Modelling the galaxy as an infinitely thin rotating disk (tilted ring model) with an exponential intensity profile and simulating the spectroscopic observation process produces a \textit{synthetic} rotation curve. Matching the synthetic RC thus obtained with the actual observed rotation curve allows to infer from the best-fitting simulated RC the true intrinsic rotation-curve parameters. \\
As input parameters for the whole simulation process, we use the redshift of the galaxy, the FWHM (Full Width at Half Maximum) of the seeing disk, the photometric scale length $r_d$, the inclination angle $i$ and the position angle $\theta$ of the galaxy. We obtain the photometric scale length $r_d$ by fitting a 2-component model with GALFIT \citep{peng02} to the corresponding HST/ACS V-band images. We assume a de Vaucouleurs profile for the bulge and an exponential profile for the disk. We stress that the rotation curve modelling explained below is robust against the assumed morphology, and changes in the morphological decomposition (like e.g. a free-n S\'{e}rsic bulge) do not affect our results. Values of the axis ratio $a/b$ ($a$ and $b$ are the apparent semi-major and semi-minor axis, respectively) and the position angle were previously determined using the SExtractor package \citep{bertin96} on the HST/ACS V-band images and are adopted from the STAGES master catalogue \citep{stagescat09}. Assuming an intrinsic disk thickness of $q=0.2$, we computed the inclination from the axis ratio \citep[e.g.][]{tully98}.\\
A robust value for the FWHM of the seeing disk is an important ingredient in correctly simulating the observation process. The presence of three reference star spectra in each quadrant of a pointing allows to determine the seeing conditions during spectroscopy. This is clearly superior to estimates inferred from the Differential Image Motion Monitor (DIMM). We determine the FWHM of the stellar spectra along the spectral axis for each MOS image. We adopt the median of the three reference stars as our input parameter for the simulation. Seeing values increase towards shorter wavelengths and a standard equation to describe this variation is given by \citet{sarazin03}: 
\begin{equation}
\mathrm{FWHM(\lambda)}\cong \mathrm{FWHM(\lambda_c)} \left( \frac{\lambda_c}{\lambda}\right)^{1/5},
\end{equation}
where we use $\lambda_c=5100\mathring{A}$ as the central wavelength of the HR-blue grism. We obtain seeing values between $0\farcs60$ and $1\farcs27$ with an average of $0\farcs93$.\\
In the following, we describe the simulation process in more detail. We assume an intrinsic rotation function of the following form \citep{courteau97}:
\begin{equation}
V_{\mathrm{rot}}^{\mathrm{int}}(r)=\frac{V_{\mathrm{max}}r}{\left(r^{\alpha}+r_{t}^{\alpha}\right)^{1/\alpha}}.
\label{eq:rc}
\end{equation}
At the turn-over radius $r_t$ the linear rise for small radii (rigid-body region) turns over into a flat region with asymptotic velocity $V_{\mathrm{max}}$, where the dark matter halo dominates the mass distribution. The factor $\alpha$ determines the sharpness of the turnover and is set to $\alpha = 5$ \citep{boehm04}. Hence, $V_{max}$ and $r_t$ are the two remaining free parameters in the fitting process.\\
As a starting point for the simulation process, we generate a two-dimensional (2D) exponential intensity profile on a pixel grid: 
\begin{equation}
      I(x,y) = I_{0}\;e^{-\frac{r}{r_d}},
      \label{eq:rd}
\end{equation}
where $r=\sqrt{x^{2}+y^{2}}$ and $I_0$ is the intensity at the galactic centre at the pixel position $(x,y)=(0,0)$. Then we superimpose (on this intensity profile) a 2D velocity field using the above rotational law (Eq. \ref{eq:rc}). This results in a rotating disk seen face-on. In the next step, we obtain the correct geometric projection. To this end, we rotate our galaxy, i.e. the 2D intensity and velocity field, according to the inclination angle $i$ (rotation around the y-axis) and the position angle $\theta$ (rotation around the line-of-sight or z-axis). For the velocity field we calculate the line-of-sight component only, since this is the component that is responsible for the Doppler shift and hence measured in a spectrum. We show an example of such a projected 2D intensity and velocity field in Fig. \ref{fig:sim1} (left panel). \\
Besides geometric projection effects we also have to consider the blurring effect due to seeing. We take this into account by convolving the intensity and velocity field with a Gaussian point spread function (PSF) (the FWHM is obtained from the seeing value during spectroscopy).
Then, as in the observations, we place a slit along the apparent major axis of the galaxy. For the following slit integration the velocity field is weighted by the normalised intensity profile, since velocities stemming from brighter regions contribute more to the signal in a spectrum. Prior to this intensity weighting of the velocity field, we shift the intensity profile according to the mismatch between the kinematic and luminous centre, which we determined in the course of rotation-curve extraction (see Sect. \ref{sec:rce}). Then, for each pixel position along the spatial axis (x-axis) the velocity values within the slit borders are integrated along the direction of dispersion (y-direction). Note that the integration is in general not perpendicular to the slit direction since most MOS slits are tilted. Effectively, we calculate an intensity-weighted average of the velocity values inside the slit at each spatial position. For the integration area at each spatial position we account for the size of the boxcar filter (3 or 5 pixels) that was used for the extraction of the rotation curve. Thus, we obtain a simulated value for the rotation velocity at the position of each pixel along the spatial axis. This position-velocity information is the end product of our simulation and defines a \textit{synthetic} rotation curve. The right panel of Fig. \ref{fig:sim1} shows a scheme of the integration process of the PSF-convolved and intensity-weighted velocity field. Unlike shown in the figure, we actually oversample the pixel grid by a typical factor of five to improve the precision of the computation. 

\subsection{Rotation-curve fitting}
\label{sec:rcf}
\begin{figure}[]
   \centering
   \includegraphics[angle=0,width=\columnwidth]{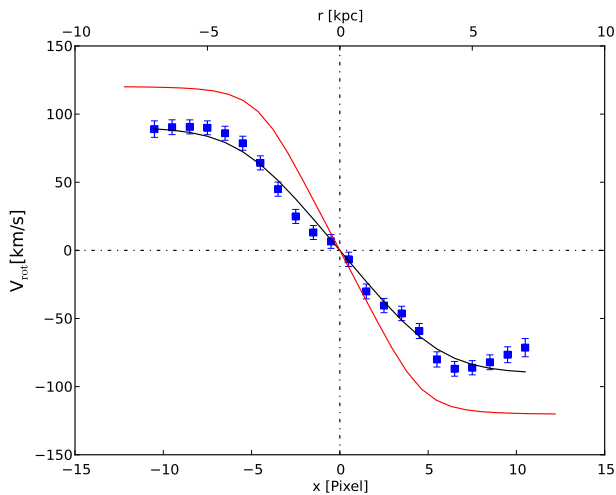}
      \caption{Example of rotation-curve fitting: The blue squares indicate the observed rotation velocities as a function of the position along the spatial axis of the spectrum (in pix). The black line depicts the best-fit synthetic RC to these data points. From the best-fit parameters (here $V_{\mathrm{max}}=118 \mathrm{km/s}$, $r_t=4.1$ pix) we can infer the intrinsic shape of the rotation curve (in red). Here the x-axis is the galactocentric radius (in kpc) (here $\theta=-36\fdg5$, $i=68\fdg5$, $z=0.17$, $\mathrm{FWHM}=0\farcs7$). Geometric and observational effects lower the observed rotation velocities.  
              }
         \label{fig:sim2}
   \end{figure}

For each galaxy we vary the free parameters ($V_{\mathrm{max}}$, $r_t$) of the synthetic RC such that it best fits the observed RC. We do this by minimizing the error-weighted $\chi^{2}$:
\begin{equation}
 \chi^{2}=\sum_{i}\left(\frac{V_{\mathrm{syn}}(x_i)-V_{\mathrm{obs}}(x_i)}{V_{\mathrm{err}}(x_i)}\right) ^{2}, 
 \label{eq:rcf}
\end{equation} 
 where the $x_i$ are the spatial positions of the observed and the simulated RC with velocities $V_{\mathrm{obs}}(x_i)$ and uncertainties $V_{\mathrm{err}}(x_i)$. Similar to \citet{pizagno07} we add $5$ $\mathrm{kms^{-1}}$ in quadrature to all velocity uncertainties to avoid the fit being dominated by high S/N data points. This is motivated by non-circular motions of the same magnitude, present in all disk galaxies. %We adopt a python implementation of a Sequential Least SQuares Programming (SLSQP) optimisation, originally encoded by \citet{kraft88}. This method has the advantage to optionally add constraining conditions or equations to the minimisation process. 
 We use the photometric scale length $r_d$ and a typical value of $100\mathrm{km/s}$ as initial guesses for the free parameters $r_t$ and $V_\mathrm{max}$. Fig. \ref{fig:sim2} gives an example of a best-fitting synthetic RC and the inferred intrinsic rotation velocities. Note that while the observed data points and the best-fitting RC are plotted against the pixel position along the spatial axis (i.e. perpendicular to the direction of dispersion), the intrinsic RC shows the intrinsic rotational velocities (see Eq. \ref{eq:rc}) as a function of galacto-centric radius.\\
Rotation curves that lack data points in the flat region tend to result in overestimated turnover radii and $V_{\mathrm{max}}$-values. To tackle this issue we constrain $r_t$ in the minimisation process (a python implementation of a Sequential Least SQuares Programming (SLSQP) optimisation, originally by \citet{kraft88}). We allow the best-fit $r_t$-values to vary only within a $\sigma_{t}$-interval around the initial value $r_d$. We determine the scatter $\sigma_{t}$ solely using rotation curves that result in best-fit $V_{\mathrm{max}}$-values changing by less than $5\%$ when fitted with or without $r_t$ as a free parameter. $\sigma_{t}$ is then set to be the standard deviation of the relative differences between $r_d$ (i.e. $r_t$ held fixed) and $r_t$ (i.e. $r_t$ free parameter), multiplied by $r_d$. We obtain a value of $\sigma_{t}=0.176*r_d$. \\
We estimate the errors of the best-fit parameters $V_{\mathrm{max}}$ and $r_t$ by generating bootstrap samples of a given observed rotation curve and calculating the standard deviation for the distribution of the best-fit parameters obtained from each iteration. To this end, for each data point (at a certain position $x_i$) the velocity value $V_{\mathrm{obs}}(x_i)$ is drawn from a Gaussian distribution with mean value $V_\mathrm{obs}(x_i)$ and dispersion $V_{\mathrm{err}}(x_i)$. Additionally, in each bootstrap iteration two data points are randomly excluded from the fitting process to account for a possible strong influence of single values on the best-fit parameters. Furthermore, the inclination and position angle are allowed to vary for each bootstrap iteration according to a normal distribution around the preset value (with a scatter $\sigma=2^{\circ}$). This accounts for the uncertainty in the derivation of the inclination angle and a possible slight misalignment between slit direction and apparent major axis. \\

\subsection{Rotation-curve quality assessment}
\begin{figure}[]
   \centering
   \includegraphics[angle=0,width=\columnwidth]{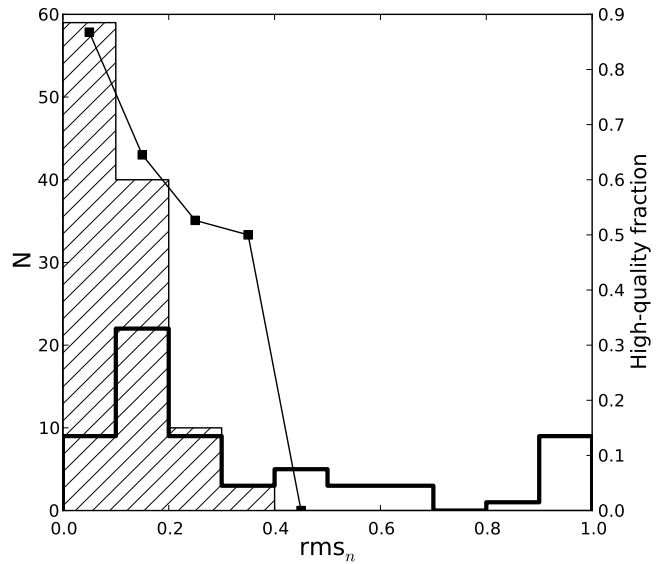}
      \caption{The histograms show the $\mathrm{rms}_{n}$ distribution for the low- (thick line) and high-quality (thin line, hatched) TF objects of our whole sample. For each histogram bin, the relative fraction of high-quality objects is depicted (black squares). It is well correlated with the $\mathrm{rms}_{n}$ value of the RC fit.}
         \label{fig:rms}
\end{figure}

Prior to any RC fitting, we inspect each RC visually and classify it according to the following scheme: non-rotating, low-quality Tully-Fisher object, high-quality Tully-Fisher object. The non-rotating class holds galaxies for which a clear distinction between an approaching and receding side is not feasible. Only five galaxies fall into this class (all of them cluster members). We exclude these objects from further analysis. Tully-Fisher objects (176 in total) show reasonable disk rotation. We assign galaxies to the low-quality class when they have a high degree of distortion, an RC with small extent or rigid-body rotation.\\
A more quantitative measure for the suitability of a galaxy as a TF object is provided by the (error-weighted) root mean square (rms) of the RC fit, i.e. the normalised sum of the squared differences between the data points and the best-fit curve. Rotation curves deviating from a Courteau model (see Eq. \ref{eq:rc}) will inevitably produce higher rms values than undistorted ones. Since absolute rms values by construction are higher for fast rotators, a normalised $\mathrm{rms}_{n}=\mathrm{rms}/V_\mathrm{max}$ is a more reliable measure. Of course, symmetric RCs that do not sufficiently trace the flat region or are identified as rigid body rotators, might also result in low rms values, despite being classified as low-quality objects. Hence, the correlation between the visual classification and the $\mathrm{rms}_{n}$ measure is not expected to be one-to-one. Fig. \ref{fig:rms} plots the fraction of high-quality objects as a function of the $\mathrm{rms}_{n}$ values. Both measures are well correlated and result in a highly significant ($p<0.01\%$) Spearman rank-correlation of $\rho=-0.52$.\\
Table \ref{table:rms} gives an overview of the mean $\mathrm{rms}_{n}$ values for various subsamples and confirms that high-quality objects have lower $\mathrm{rms}_{n}$ values than low-quality objects. This indicates that the latter on average have more distorted kinematics, i.e. their deviation from an ideal Courteau model is stronger. This higher degree of kinematic distortion is also reflected in the mean values of the asymmetry measures defined in Paper I. There, we introduced an asymmetry index $A$ following \citet{dale01}, which compares corresponding rotation velocities between the approaching and receding side of a rotation curve, and a visual asymmetry parameter $A_{\mathrm{visual2}}$, which distinguishes between three classes: undistorted (0), slightly distorted (1) and heavily distorted (2). High-quality objects ($A=14\pm1\%$, $A_{\mathrm{visual2}}=0.51\pm0.06$) have significantly lower values than low-quality objects ($A=49\pm6\%$, $A_{\mathrm{visual2}}=1.10\pm0.10$) in both classification schemes.

\begin{table}    % is used to refer this table in the text
\centering                          % used for centering table
\ra{1.0}
\resizebox{\columnwidth}{!}{
\begin{tabular}{c c c c} \toprule       % centered columns (4 columns)
%\hline\hline                % inserts double horizontal lines
Population & Member& high-quality & $\mathrm{rms}_{n}$ \\&galaxies $N$& fraction& [$\%$] \\    % table heading 
\hline            % inserts single horizontal line
\noalign{\smallskip}   
   Total & 181 & 0.62 & $20.0\pm1.7$ \\      % inserting body of the table
\noalign{\smallskip}  
  Cluster (all) & 95 & 0.58 & $22.2\pm2.5$ \\
\noalign{\smallskip}    
   Cluster (hqTF) & 55 & 1 & $11.5\pm0.8$ \\
\noalign{\smallskip}    
   Cluster (lqTF) & 35 & 0 & $34.7\pm4.9$\\
%\noalign{\smallskip}    
 %  Cluster (nonTF) & 5 & -1 & $52.6\pm19.7$\\
\noalign{\smallskip}  
  Field (all) & 86 & 0.66 & $19.8\pm2.5$\\
\noalign{\smallskip}    
   Field (hqTF) & 57 & 1 & $11.4\pm1.1$ \\
\noalign{\smallskip}    
   Field (lqTF) & 29 & 0 & $36.4\pm6.3$ \\
\noalign{\smallskip}  
\bottomrule                                  %inserts single line
\end{tabular}}
\caption{Mean $\mathrm{rms}_{n}$ values from RC fitting for our sample, as well as for certain subsets. Cluster and field galaxies are further subdivided into a low and a high-quality TF class. Five cluster objects are classified as non-rotating and not used in the TF analysis.}             % title of Table
\label{table:rms} 
\end{table}

\subsection{Linear regression}
Various methods are used in the literature for fitting a regression line to data points. We obtained a TF relation of the form:
\begin{equation}
M = a*\log V + b
\label{eq:tfr}
\end{equation}
by following \citet{pizagno07} and adopting a maximum likelihood (ML) method to determine the slope $a$ and intercept $b$ \citep[see also][]{weiner06,bamford06}. Applying Bayes' theorem allows to calculate the probability of a model given the observed data: 
\begin{equation}
P(\mathrm{model}\mid \mathrm{data}) \propto P(\mathrm{data}\mid \mathrm{model}) \times P(\mathrm{model}),
\label{eq:bt}
\end{equation}
where P(model) is the prior. Flat priors are sufficient in most cases \citep{dagostini05}, hence they drop out of the equation. In contrast to ordinary least square (OLS) methods this Bayesian approach allows to include observational errors on both axes as well as an \textit{intrinsic} scatter $\sigma_{\mathrm{int}}$. Even if measurement errors were negligible, a relationship between two physical variables will have some scatter. This intrinsic scatter reflects variations in physical parameters that are not explained by the scatter of the observables alone \citep{kelly07}. We assume Gaussian distributions for the (uncorrelated) observational errors as well as for the intrinsic scatter. We further assume a uniform intrinsic distribution of the independent variable $x_i$, so that the log-likelihood to maximise reads \citep[e.g.][]{pizagno07}:
\begin{equation}
\begin{aligned}
\ln L_{U}(a,b,\sigma_{\mathrm{int}})=&-\sum_{i}\ln \left( \sigma_{\mathrm{int}}^{2}+a^{2}\sigma^{2}_{x,i}+\sigma^{2}_{y,i}\right) \\
 &-\sum_{i} \frac{\left(y_{i}-ax_{i} - b\right)^{2} }{\sigma_{\mathrm{int}}^{2}+a^{2}\sigma^{2}_{x,i}+\sigma^{2}_{y,i}} + \mathrm{const.},
\end{aligned}
\label{eq:logL} 
\end{equation}
where $\sigma_{y,i}$ and $\sigma_{x,i}$ are the measurement errors of the dependent variable $y_i$ and independent variable $x_{i}$, respectively. Compared to traditional least square methods, this fitting approach has the advantage that data points with large errors in either of the variables will have less influence on the final result. Moreover, the inclusion of the intrinsic scatter prevents data points with very small observational errors from dominating the fit.\\
However, \citet{kelly07} emphasised that adopting a uniform distribution for the independent variable can produce a bias in the ML estimate and is not appropriate for most situations (as is e.g. the case for the distribution of the absolute B-band magnitudes of our sample, see Fig. \ref{fig:schechter}). Instead, they proposed to approximate the intrinsic distribution of the independent variable using a mixture of Gaussian functions. For simplicity, we adopt a single Gaussian with mean $\mu$ and variance $\tau^{2}$ determined from a fit to the histogram of the independent variable of our whole sample. Then, the log-likelihood of Eq. \ref{eq:logL} becomes:
\begin{equation}
\begin{aligned}
\ln L_{G}(a,b,\sigma_{\mathrm{int}}) &=-\sum_{i}\ln \left( \Theta^{2}\right) - \sum_{i} \frac{\left(y_{i} - E\right)^{2} }{\Theta^{2}} + \mathrm{const.},\\
E&=b+\frac{x_{i}a\tau^{2}}{\tau^{2}+\sigma^{2}_{x,i}} + \frac{\mu a\sigma^{2}_{x,i}}{\tau^{2}+\sigma^{2}_{x,i}},\\
\Theta^{2} &=a^{2}\tau^{2}+\sigma_{\mathrm{int}}^{2}+\sigma^{2}_{y,i}-\frac{a^{2}\tau^{4}}{\tau^{2}+\sigma^{2}_{x,i}},
\end{aligned}
\label{eq:logLK} 
\end{equation}
where $E$ and $\Theta^{2}$ are the expectation value and variance of $y_{i}$ at $x_{i}$, drawn from a Gaussian distribution with mean $\mu$ and variance $\tau^{2}$. The covariance between the observational errors $\sigma_{y,i}$ and $\sigma_{x,i}$ was set to zero. For $\tau^{2}\rightarrow\infty$, i.e. an infinitely broad Gaussian, $L_{G}$ of Eq. \ref{eq:logLK} converges to $L_{U}$ of Eq. \ref{eq:logL}.\\
 We are aware that the intrinsic distribution of luminosities would rather be a Schechter-like function. However, a Gaussian is a good approximation to the observed luminosity function of spiral galaxies only \citep{binggeli88}. Furthermore, it is in any case more realistic than a flat distribution.\\
We perform the minimisation of the (negative) log-likelihood using a downhill simplex algorithm. We estimate the errors of the best-fit parameters $a$, $b$ and $\sigma_{\mathrm{int}}$ utilising bootstrap simulations ($n=1000$). \\
Instead of fitting the ``forward" TFR of Eq. \ref{eq:tfr}, i.e. $y=M$ and $x=\log V$, many studies use the ``inverse" relation: 
\begin{equation}
\log V = \frac{1}{a}M - \frac{b}{a}
\label{eq:tfri}
\end{equation}
The inverse fit, i.e. the velocity treated as the dependent variable, is less prone to incompleteness biases that arise from the magnitude limit of a sample \citep{willick94,tully12}. \citet{kelly07} demonstrated that if the sample selection is based on the independent variable of the fit (the luminosity in our case), then the best-fit parameters are not influenced by selection effects. However, treating either the luminosity (forward fit) or the velocity (inverse fit) as the dependent variable does not reflect the situation properly, since by construction in each case the scatter in the dependent variable is minimised. To study the underlying relationship between two observables a method treating them symmetrically could be more appropriate. Such methods utilise both the forward and the inverse best-fit parameters to calculate a best-fit slope that lies between the two extremes. \citet{isobe90} describe for instance bisector regression, orthogonal regression or reduced major-axis regression. \citet{tully12} simulated the selection bias for different fitting methods and recommend the inverse relation when using the TFR as a means of distance measurement. On the other hand, to study the physical impacts on the TFR a bivariate fit would be a good compromise.\\
We choose to use the inverse fit when comparing different populations or distant to local studies. However, for the investigation of Tully-Fisher residuals (luminosity offsets) we use the bisector parameters calculated from the forward and inverse fits as in \citet{isobe90}. Whatever the type of fit, we always report the resulting parameters in terms of a forward relation.

\subsection{Incompleteness bias correction}
\begin{figure}[]
   \centering
   \includegraphics[angle=0,width=\columnwidth]{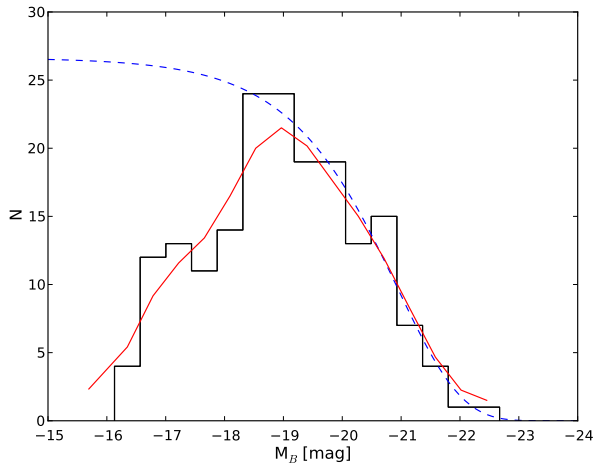}
      \caption{B-band luminosity histogram of our spectroscopic sample. The red solid line depicts a Hanning-smoothed fit to this histogram. The blue dashed line indicates a Schechter fit to the bright end ($M_{B}\leq -19\fm5$). The obtained parameter is $M^{*}=-20\fm9$ for the characteristic luminosity. We assume a slope at the faint end of $\alpha = -1.00$. The incompleteness of our sample is given by the difference between the two lines.}
         \label{fig:schechter}
   \end{figure}

\begin{figure}[]
   \centering
   \includegraphics[angle=0,width=\columnwidth]{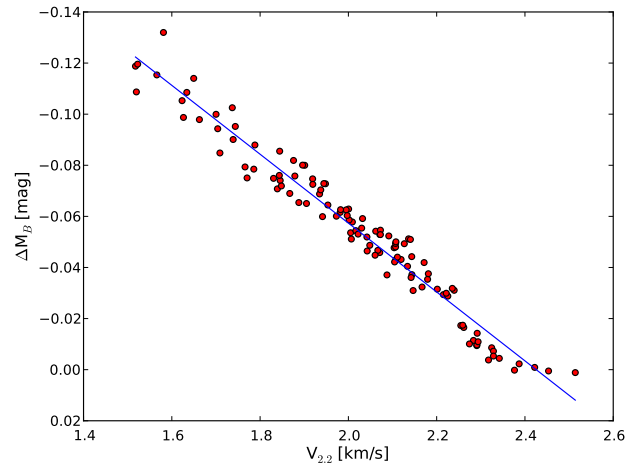}
      \caption{Simulated incompleteness bias obtained from Monte-Carlo simulations. The data points depict the average TF residual for a certain velocity. The blue solid line indicates a linear fit to these data points.  }
         \label{fig:icbias}
   \end{figure}

By construction magnitude-limited samples result in an incomplete coverage of the faint end of the luminosity function. In a TF sample with magnitude limit faint slow rotators are under-represented. In the end, this leads to an underestimation of the TFR slope accompanied by an overestimation of the intercept.\\
\citet{giovanelli97} developed a method to account for this bias. First of all, we have to assume an intrinsic luminosity distribution according to a Schechter luminosity function. Fig. \ref{fig:schechter} shows a histogram of the B-band luminosities and a Hanning-smoothed fit to this histogram. The latter is then compared to a fit of a Schechter luminosity function to the bright end of the data ($M_{B}\leq -19\fm5$). For these luminosities, the sample is assumed to be complete. The ratio between the two lines (maximum set to 1) provides us with a measure $c(M_B)$ for the completeness of the sample. \\
To simulate the incompleteness bias we generate Monte-Carlo simulations with $10^{4}$ iterations. For a given rotation velocity a luminosity is calculated according to an intrinsic TFR. The TFR scatter is simulated by randomly drawing from a normal distribution with a dispersion equal to the scatter in our data. The luminosity value $M_B$ thus obtained now has the probability $c(M_B)$ to enter the sample. Of course, fainter galaxies have a lower chance to be considered. The average residual from the TFR gives an estimate of the magnitude bias for a given rotational velocity (see Fig. \ref{fig:icbias}). The magnitude bias $\Delta M_{B} = f(V_{2.2})$ is then approximated by a linear fit. To account for the slightly different luminosity and redshift distributions of the field and cluster sample, the magnitude bias was computed for the two populations separately.

\section{Tully-Fisher relations}
Quantitative comparisons between different Tully-Fisher studies are very difficult. First of all, there can be differences in the observational approach. The rotation of a galaxy may be traced e.g. by radio data of the \ion{H}{I} disk or by optical emission lines stemming from \ion{H}{II} regions. Radio observations have the advantage of tracing the rotation of a galaxy to large radii, where the dark matter halo dominates the mass density \citep[e.g.][]{verheijen01}. \citet{pizagno07} pointed out that obtaining velocity-widths from \ion{H}{I} rotation curves results in rotation velocities systematically higher for fainter galaxies compared to optical observations. This leads to a shallower TFR. Furthermore, some studies use integrated line-width while others exploit spatially resolved rotation curves or even velocity fields. Compared to the traditional long-slit spectroscopy, IFU (Integral Field Unit) velocity fields might be superior in identifying galaxies with distorted and peculiar kinematics. Besides this, differences in sample selection (magnitude limit, morphological type, constraints on kinematic distortion), different fitting methods and considerations of observational effects are the most common sources of discrepancies. To avoid a possible bias, we focus on the comparison of subsamples in our analysis, e.g. between field and cluster galaxies or the blue cloud and dusty red population. \\
Studies designed to use the TFR as a distance indicator (usually with the goal of constraining the Hubble constant $H_0$), need a pruned sample containing only ``model" galaxies such as undistorted late-type spirals. This guarantees a tight TFR with a small scatter. However, we also intend to investigate galaxies that deviate from this ideal, which might reveal information on interaction mechanisms in the cluster environment.

\subsection{Basic fits} 
\begin{figure}[]
   \centering
   \includegraphics[angle=0,width=\columnwidth]{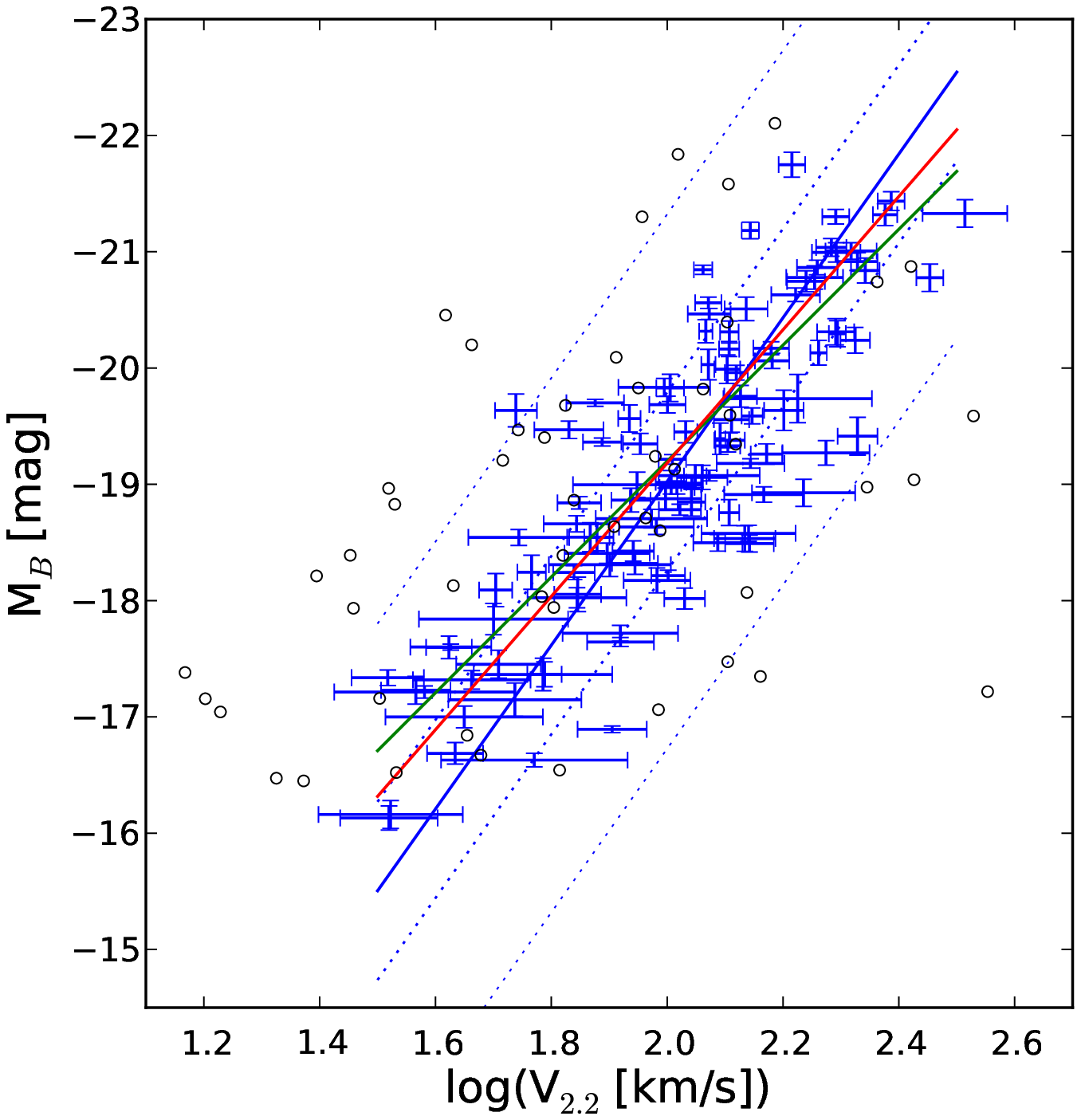}
      \caption{B-band TFR for the combined sample of cluster (55) and field (57) galaxies. Only high-quality objects are considered. The blue solid line indicates the best-fit inverse regression line. The blue dotted lines depict the corresponding $1\sigma$ and $3\sigma$ intervals, respectively. The green and red line are the best-fit forward and bisector fits, respectively. Black open circles indicate the positions of the low-quality objects.}
         \label{fig:BTFRcf}
   \end{figure}

Table \ref{tab:fitcomp} gives an overview of the parameter estimates using the maximum likelihood method with a Gaussian intrinsic distribution of the independent variable (Eq. \ref{eq:logLK}). Note that using a uniform intrinsic distribution instead (Eq. \ref{eq:logL}, not shown in the table) yields consistent values within the $1\sigma$ errors.\\
 First, we concentrate on the B-band TFR using only rotation curves of our high quality class.
\begin{table*}\centering
\ra{1.2}
\resizebox{\textwidth}{!}{
\begin{tabular}{@{}rrrrcrrr@{}}\toprule
& \multicolumn{3}{c}{B-band TFR}&\phantom{abc}&\multicolumn{3}{c}{Stellar-mass TFR}\\
\cmidrule{2-4} \cmidrule{6-8}
& slope $a$ & intercept $b$ & scatter ($\sigma_{\mathrm{int}}$) && slope $a$ & intercept $b$ & scatter ($\sigma_{\mathrm{int}}$)        \\ \midrule
Total (hq, 112 galaxies)\\
forward & $-5.10\pm0.30$ & $-8.87\pm0.61$ & $0.66\pm0.05$ && $2.41\pm0.12$ & $4.95\pm0.24$ & $0.26\pm0.02$\\
inverse & $-7.04\pm0.46$ & $-4.95\pm0.95$ & $0.77\pm0.10$ && $3.04\pm0.19$ & $3.69\pm0.39$ & $0.29\pm0.03$\\
bisector& $-5.92\pm0.38$ & $-7.20\pm0.77$ & $0.71\pm0.07$ && $2.69\pm0.15$ & $4.40\pm0.31$ & $0.27\pm0.03$\\
Cluster (hq, 55 galaxies) \\
forward & $-5.28\pm0.32$ & $-8.33\pm0.64$ & $0.57\pm0.07$ && $2.34\pm0.16$ & $5.14\pm0.33$ & $0.25\pm0.02$\\
inverse & $-6.58\pm0.61$ & $-5.69\pm1.25$ & $0.64\pm0.10$ && $2.86\pm0.21$ & $4.07\pm0.42$ & $0.27\pm0.03$\\
bisector& $-5.86\pm0.45$ & $-7.12\pm0.93$ & $0.60\pm0.08$ && $2.57\pm0.18$ & $4.65\pm0.38$ & $0.26\pm0.03$\\
Field (hq, 57 galaxies)\\
forward & $-5.03\pm0.44$ & $-9.15\pm0.89$ & $0.69\pm0.07$ && $2.48\pm0.16$ & $4.81\pm0.34$ & $0.26\pm0.04$\\
inverse & $-7.51\pm0.81$ & $-4.20\pm1.66$ & $0.84\pm0.15$ && $3.25\pm0.36$ & $3.27\pm0.75$ & $0.30\pm0.06$\\
bisector& $-6.03\pm0.61$ & $-7.17\pm1.22$ & $0.76\pm0.11$ && $2.82\pm0.25$ & $4.14\pm0.50$ & $0.28\pm0.05$\\
Total (h+lq, 176 galaxies)\\
forward & $-4.09\pm0.38$ & $-10.85\pm0.74$ & $0.90\pm0.08$ && $1.94\pm0.16$ & $5.95\pm0.32$ & $0.43\pm0.04$\\
inverse & $-7.08\pm0.64$ & $-5.03\pm1.30$ & $1.20\pm0.17$ &&  $3.37\pm0.35$ & $3.15\pm0.70$ & $0.58\pm0.09$\\
bisector& $-5.20\pm0.50$ & $-8.96\pm0.97$ & $1.04\pm0.12$ && $2.49\pm0.25$ & $5.01\pm0.48$ & $0.50\pm0.06$\\
\bottomrule
\end{tabular}}
\caption{Parameter estimates using the maximum likelihood method with a Gaussian intrinsic distribution of the independent variable. For each subpopulation forward, inverse and bisector parameters are listed. All values are given in the forward representation of the Tully-Fisher relation.}
\label{tab:fitcomp}
\end{table*}
Inverse fits to the sample of field (57 objects, $a=-7.51\pm0.81$) and cluster (55 objects, $a=-6.58\pm0.61$) galaxies yield consistent slopes within the $1\sigma$-errors. Hence, in the following subsections we mainly use a fixed slope, determined from an inverse fit to the combined sample of cluster and field galaxies. This is a common approach to reduce the number of free parameters and is justified by the fact that most other TF studies also do not find significant differences in the slope between cluster and field galaxies \citep{bamford06,moran07,nakamura06,ziegler03}. Fig. \ref{fig:BTFRcf} shows an inverse fit to the high-quality objects of our whole sample. As a comparison we also depict the (shallower) slopes of the bisector and forward fit as well as the position of the low-quality galaxies. \\
Note that an inverse fit minimises the scatter with respect to the rotation velocity and not to the luminosity. Hence, an intrinsic scatter obtained from an inverse fit and transformed into a value with respect to the luminosity is always larger than if directly obtained using a forward fit. This needs to be considered when comparing to studies that adopt a forward fit. Furthermore, reducing the number of free parameters (e.g. holding the slope fixed) can also slightly change the best-fit intrinsic scatter.\\
For the stellar-mass TFR the parameters of field and cluster galaxies also agree within the error bars.\\
Including also the low-quality objects in the fit (176 objects in total) increases the intrinsic scatter significantly (see Table \ref{tab:fitcomp}). Interestingly, the best-fit (inverse) values for slope and intercept barely change. This is mainly due to the weighting capability of our linear regression method. The adopted bootstrapping method for RC fitting (see end of Section \ref{sec:rcf}) assigns larger errors to low-quality objects ($\delta V = 19\pm 3$ km/s) compared to high-quality objects ($\delta V = 10\pm 1$ km/s).\\
 For fixed slopes ($a=-7.04$ and $a=3.04$ for the B-band and stellar-mass TFR) we find a difference of $\Delta M_{B}=-0\fm70\pm0\fm28$ and $\Delta M_{\ast}=0.46\pm0.16$ when comparing the TF intercept between low- and high-quality objects individually, i.e. distorted objects are preferentially located above the TFR of high-quality objects. This means they are on average too luminous/stellar-massive for a given velocity or have too low a velocity for a given luminosity/stellar mass. The comparable shifts ($\Delta V_{2.2}\sim 0.12\pm0.5$, when transformed into a velocity shift) for the stellar-mass and the B-band TFR as well as a similar increase in the intrinsic scatter ($\sigma_{\mathrm{int}}\sim 0.18\pm0.3$ in velocity) indicate that mainly an underestimated rotation velocity causes this shift for low-quality objects.
 %Concerning this matter \citet{weiner06} introduced a new kinematic measure $S_{0.5}$ that also accounts for non-circular gas motions through an additional velocity dispersion term. Eventually this may lead to a tighter TFR for kinematically distorted galaxies \citep[e.g.][]{kassin07}.

\subsection{Cluster vs. field relation}
\begin{figure*}[]
   \centering
   \includegraphics[angle=0,width=\textwidth]{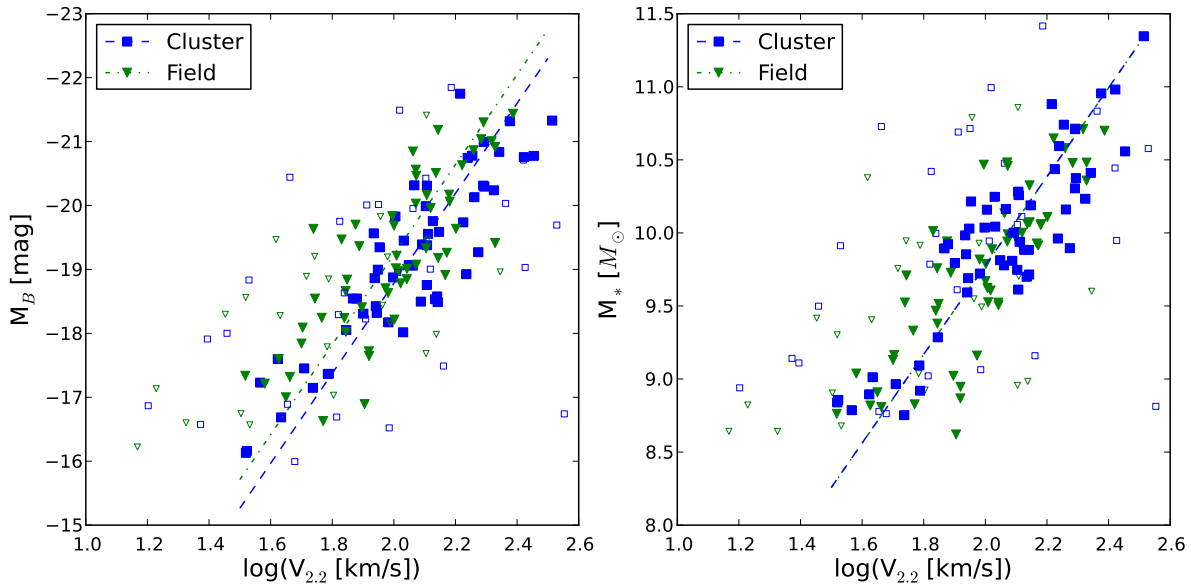}
      \caption{TFR for the sample of high-quality cluster (55 objects; blue squares) and field (57 objects; green triangles) galaxies. Open symbols indicate the corresponding low quality objects. The blue dashed and green dot-dashed lines indicate the best-fit inverse regression lines of cluster and field galaxies, respectively. B-band TFR (left panel): Field galaxies are brighter by $\Delta M_{B}=0\fm42\pm0\fm15$. SmTFR (right panel): We find no difference between the intercept and intrinsic scatter of the two populations ($\Delta M_{\ast}=0.00\pm0.07$, $\Delta \sigma_{\mathrm{int}}=0.01\pm0.04$).}
         \label{fig:BTFRcfc}
   \end{figure*}

Holding the TFR slope fixed at $a=-7.04$, we now compare the B-band TFR of high-quality cluster and field galaxies. We find an offset between the two populations (Fig. \ref{fig:BTFRcfc}, left panel) with field galaxies being slightly brighter than their cluster counterparts ($\Delta M_{B}=-0\fm42\pm0\fm15$).\\
 Furthermore, we find a slight difference between the mean SED types. Field galaxies have on average a bluer SED than cluster galaxies ($\langle T_{\mathrm{type}} \rangle$=5.8 vs. 5.2 $\pm0.2$), which could explain part of the offset in the TFR. T-types of 5 and 6 correspond to typical spectra of Sc and Scd spirals, respectively. A linear fit to the relation of the B-band residuals as a function of the SED-type yields a brightening of $\Delta M_{B}\sim-0\fm09 \pm0\fm05$ for an SED-type change of $\Delta T_{\mathrm{type}}=1$.\\
 Another aspect might be the deviating redshift distribution of both samples. According to a trend seen in previous TF-evolution studies \citep[e.g.][]{ziegler02,weiner06,boehm04,bamford06,miller11,fernando10}, the slightly higher redshift of our field galaxies ($\langle z \rangle$=0.245 vs. 0.165) coincides with higher star-formation rates \citep[e.g.][]{hopkins04} and consequently higher B-band luminosities. Most studies measure a brightening in the B-band consistent with $\Delta M_{B}=-1\fm0\pm0\fm5$ from $z=0$ to $z\sim1$. Galaxy formation models based on $\Lambda$CDM cosmology \citep{dutton11a} produced similar values ($\Delta M_{B}=-0\fm9$). Hence we would expect an offset of $\Delta M_{B}\sim-0\fm1\pm0\fm1$ stemming from the redshift difference between our cluster and field sample.\\
Altogether, systematic differences in the composition of the cluster and field sample may account for half of the observed offset in the B-band intercept. The cluster environment associated with a quenching of star-formation could be responsible for the remaining difference ($\Delta M_{B}\sim-0\fm23 \pm0\fm19$). Of course this difference is barely significant. Furthermore, if the assumption of a fixed slope between cluster and field galaxies is not fulfilled, i.e. cluster galaxies have an intrinsically shallower slope, then an underestimation of the TF intercept is expected, when fixing the slope.\\
 On the other hand, under the assumption of such a luminosity-redshift correlation, the broader redshift distribution of the field sample ($\sigma_{z}$=0.048 vs. 0.005 for the cluster, which is a factor of $\sim$10) should result in a slightly larger TF scatter.
  The intrinsic scatter amounts to $\sigma_{\mathrm{int}}=0.68\pm0.06$ for the cluster and $\sigma_{\mathrm{int}}=0.79\pm0.08$ for the field population, which is in agreement within the errors. Given the redshift distribution of our field sample and assuming an evolution of $\Delta M_{B}=(-1\fm0\pm0\fm5)\times z$ \citep[e.g.][]{bamford06}, simulations yield an estimate of $\Delta \sigma=0.05\pm0.02$ for the increase of the scatter due to the broader redshift distribution.\\
 However, the situation changes when only low-quality Tully-Fisher objects are considered. While the difference in the intercept is similar ($\Delta M_{B}=-0\fm62\pm0\fm48$), the intrinsic scatter for the cluster sample (35 objects; $\sigma_{\mathrm{int}}=2.20\pm0.32$) is significantly higher than for the corresponding field subsample (29 objects; $\sigma_{\mathrm{int}}=1.42\pm0.25$). Whereas cluster galaxies with more regular kinematics show no enhanced scatter, cluster-specific interaction processes seem to induce a larger TF scatter in galaxies with distorted kinematics.\\
However, when we examine the high-quality stellar-mass TFR, the differences between cluster and field galaxies disappear. Again, we fix the slope to the (inverse) value of the combined sample ($a=3.04$). Then, the intercept ($\Delta M_{\ast}=0.00\pm0.07$) and intrinsic scatter ($\Delta \sigma_{\mathrm{int}}=0.01\pm0.04$) for both populations are in total agreement. Fig. \ref{fig:BTFRcfc} (right panel) illustrates the coinciding sample of cluster and field galaxies for the stellar-mass TFR.\\
This finding suggests that any measured differences between the TFR of cluster and field galaxies stem from different M/L ratios and hence star-formation rates affecting the B-band luminosity. It confirms that differences in luminosity and not in velocity induce the offset of the B-band TFR between cluster and field galaxies.

\subsection{Local comparison samples}
\begin{figure*}[]
   \centering
   \includegraphics[angle=0,width=\textwidth]{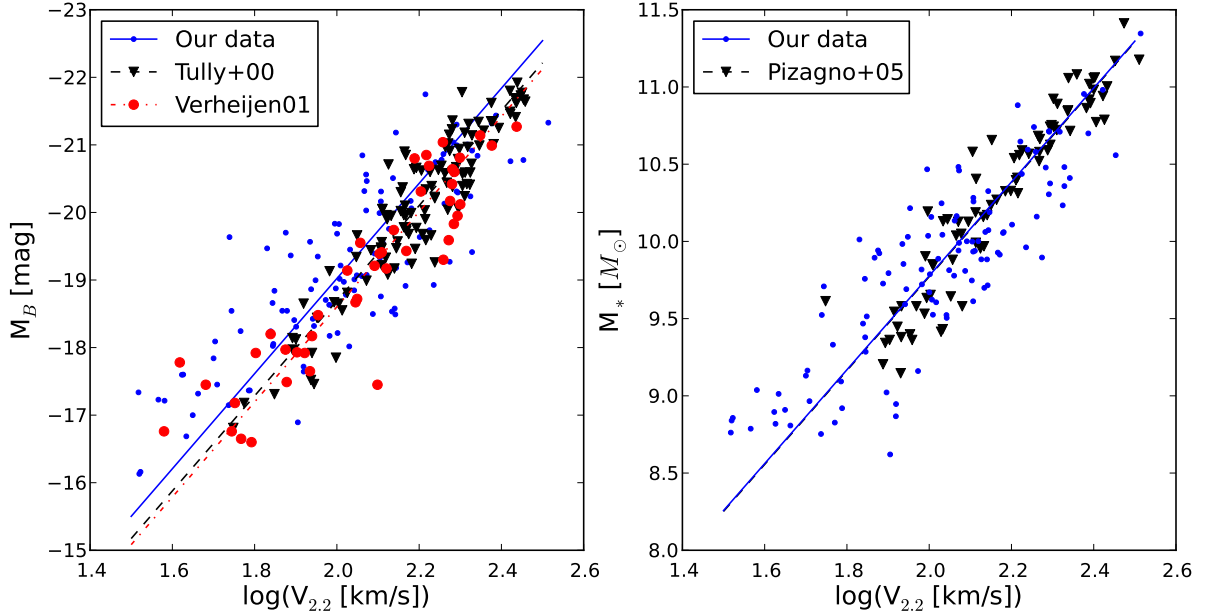}
      \caption{TF data for our total sample (blue points, solid line) and the local comparison samples. The depicted regression lines are for the fixed slope of our sample. B-band (left panel): the intercept differences are $\Delta M_{B}=0\fm33\pm0\fm10$ for the \citet[][black triangles, dashed line]{tully00} sample and $\Delta M_{B}=0\fm43\pm0\fm13$ for the sample of \citet[][red circles, dot-dashed line]{verheijen01}. Stellar mass (right panel): the intercept of our sample agrees with the local sample of \citet[][black triangles, dashed line]{pizagno05} ($\Delta M_{\ast}=0.01\pm0.04$). }
         \label{fig:TFRcomp}
   \end{figure*}

In the following we want to compare our sample to local TF studies. We select the B-band data of \citet[][Sample T hereafter]{tully00} and \citet[][Sample V hereafter]{verheijen01} as well as the stellar-mass data of \citet[][Sample P hereafter]{pizagno05} as comparison samples. Both B-band studies exploited \ion{H}{I} linewidth measurements $W_{20}$ (at $20\%$ of the peak flux) to derive rotation velocities, and applied the same intrinsic mass-dependent absorption correction as we do. The \citet{tully00} sample comprises 115 spirals (Sa-Sd) in various environments and was mainly designed to calibrate the TFR for subsequent $H_0$ determination. We also use the total sample of 45 galaxies from \citet{verheijen01} as a less pruned comparison sample, including objects with potentially distorted kinematics. \citet{pizagno05} selected 81 disk-dominated galaxies from the SDSS and derived the rotation velocity at a radius of 2.2 disk scale lengths, $V_{2.2}$, from \ion{H}{$\alpha$} rotation curves.\\
 By applying the same fitting procedure to our sample and the local samples, we rule out a potential source of discrepancies. Table \ref{tab:local} lists the TF parameters we obtain when performing a free inverse fit and when fixing the slope to a=-7.04 (B-band TFR) or a=3.04 (smTFR):
 \begin{table}\centering
\ra{1.1}
\resizebox{\columnwidth}{!}{
\begin{tabular}{@{}rrrr@{}}\toprule
& \multicolumn{3}{c}{Local TFR}\\
\cmidrule{2-4}
& slope $a$ & intercept $b$ & scatter ($\sigma_{\mathrm{int}}$) \\ \midrule
Sample T &B-band TFR\\
 & $-7.93\pm0.29$ & $-2.66\pm0.64$ & $0.43\pm0.04$\\
 & $-7.04$ fixed & $-4.62\pm0.04$ & $0.40\pm0.03$\\
Sample V & B-band TFR\\
 & $-6.98\pm0.52$ & $-4.65\pm1.11$ & $0.61\pm0.10$\\
 & $-7.04$ fixed & $-4.52\pm0.09$ & $0.61\pm0.09$\\
Sample P & stellar-mass TFR\\
 & $3.40\pm0.11$ & $2.89\pm0.23$ & $0.17\pm0.02$\\
 & $3.04$ fixed & $3.70\pm0.02$ & $0.15\pm0.02$\\
\bottomrule
\end{tabular}}
\caption{Parameter estimates of local comparison samples using the maximum likelihood method with a Gaussian intrinsic distribution of the independent variable. For each sample we carry out an inverse fit with a free and a fixed slope.}
\label{tab:local}
\end{table}
Fig. \ref{fig:TFRcomp} shows an overview of the data. Both local B-band samples (left panel) have a fainter intercept than our sample. Fixing the slope yields differences in the intercept of $\Delta M_{B}=0\fm33\pm0\fm10$ (Sample T) and $\Delta M_{B}=0\fm43\pm0\fm13$ (Sample V), respectively. This, as well as the smaller scatter, can be most likely attributed to the higher redshift of our sample. Compared to \citet{tully00} the less strict selection constraints might also contribute to the larger scatter in our data. The best-fit slopes are in good agreement with our values, although the relation of the Sample T is slightly steeper. A lack of faint fast-rotating galaxies in our sample (especially in the cluster population) seems to be responsible for this difference. As already mentioned this could (in contrast to velocity width computed from \ion{H}{I} data) partly be due to the derivation of rotational velocities from optical emission lines \citep{pizagno07}.\\
The right panel of Fig. \ref{fig:TFRcomp} shows the stellar-mass TFR in comparison to the local data of \citet{pizagno05}. When fixing the slope, the difference in the intercept between the local P~sample and our sample is negligible ($\Delta M_{\ast}=0.01\pm0.04$). This is in agreement with previous studies that find no significant evolution of the stellar-mass TFR with redshift up to $z\lesssim2$ \citep[e.g.][]{miller11,miller12}.

\subsection{Blue cloud vs. dusty red galaxies}
\label{ss:bcdr}
\begin{figure*}[]
   \centering
   \includegraphics[angle=0,width=\textwidth]{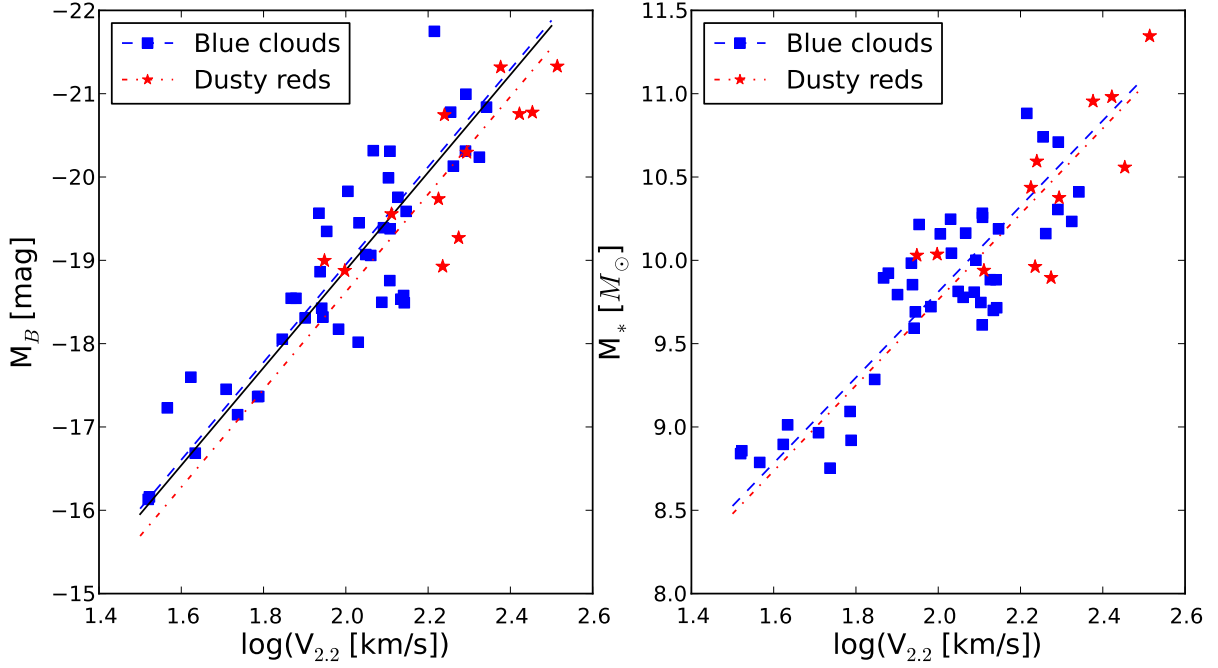}
      \caption{TFR for the sample of high-quality cluster galaxies, separated into dusty red (red stars) and blue cloud (blue squares) galaxies. The black line indicates the best-fit bisector regression line for the whole cluster sample. Fixing the slope to this value, the red dashed-dotted and blue dashed lines depict the corresponding best-fit lines for dusty reds and blue clouds, respectively. B-band (left panel): For a given velocity dusty red galaxies are on average $\Delta M_{B}=0\fm35\pm0\fm12$ fainter than blue cloud galaxies. Stellar mass (right panel): The difference between the average residuals of dusty red and blue cloud galaxies is not significant ($\Delta  M_{\ast}=0.06\pm0.05$).}
         \label{fig:TFRbcdr}
   \end{figure*}

In the U-V colour-magnitude diagram, dusty red galaxies overlap with the red sequence of ellipticals and lenticulars without star formation. This is due to significant dust extinction, similar to blue cloud galaxies, along with a low-level of star formation \citep{wolf05}. They are mainly early-type spirals and their star-formation is $\sim$four times lower than in blue cloud spirals of similar mass \citep{wolf09}. Hence, we expect that dusty red galaxies are located below the TFR of blue cloud galaxies. Fig. \ref{fig:TFRbcdr} (left panel) confirms this expectation. The dusty red galaxies preferentially populate the region below the (bisector) regression line of the whole high-quality cluster sample (blue clouds + dusty reds). The mean (error-weighted) residual amounts to $\overline{\Delta} M_{B}=0\fm24\pm0\fm11$ compared to $\overline{\Delta} M_{B}=-0\fm11\pm0\fm06$ for blue cloud galaxies, i.e. dusty red galaxies are on average $\Delta  M_{B}=0\fm35\pm0\fm12$ fainter for a given velocity than blue cloud galaxies. The trend that the cluster slope is slightly shallower might stem from the high fraction of dusty red galaxies in our sample. They populate the intermediate-to-high mass regime of our sample and lie on average below the TFR (see Fig.\ref{fig:TFRbcdr}, left panel), thus reducing the slope. Excluding the dusty red galaxies (12) from the fit yields a slope closer to the field value (43 objects, e.g. for the inverse fit $a=-7.07\pm0.74$).\\
 The right panel of Fig. \ref{fig:TFRbcdr} shows the same comparison between dusty red and blue cloud galaxies, but this time for the stellar-mass TFR. Here, the mean (error-weighted) residuals from the bisector regression line are $\overline{\Delta} M_{\ast}=-0.05\pm0.05$ for dusty red and $\overline{\Delta} M_{\ast}=0.01\pm0.02$ for blue cloud galaxies, respectively. In contrast to the B-band residuals, this is an insignificant difference ($\Delta  M_{\ast}=0.06\pm0.05$). This suggests that at given rotational velocity dusty red galaxies have similar stellar masses as blue cloud galaxies. A difference only appears in the B-band TFR as a result of different M/L values.\\
The offset in the average B-band residuals fits into the scenario that dusty red galaxies might be the progenitors of S0 galaxies. The latter are found to systematically lie below the Tully-Fisher relation of spiral galaxies \citep{aragon08}. \citet{cortesi13} investigated the kinematics of six S0 galaxies exploiting spectra of planetary nebulae. Besides a position of the disk components below the TF relation of local spirals, they also found that the bulge components populate the region above the Faber-Jackson relation of elliptical galaxies. In other words, bulges are too bright but the disks are too faint for a given velocity measure. Furthermore, the disks exhibit an increased fraction of random motion compared to spiral galaxies. The authors pointed out that neither a quiescent scenario like disk-fading nor a violent interaction like merging are likely explanations of the observations. A milder process like ram-pressure would, however, be well in compliance with what we find.\\
Our sample of 182 disk galaxies contains 14 objects (11 in the cluster; 3 in the field) with an S0 morphology. $22\%$ of the cluster galaxies with spiral morphology are dusty red galaxies, and 9 of 11 lenticulars ($82\%$) have a dusty red SED-type. Only four cluster S0 galaxies show high-quality rotation curves. Their mean (error-weighted) residual is $\overline{\Delta} M_{B}=0\fm74\pm0\fm23$ which is even fainter than the average dusty-red population. The small number statistics do not allow to draw conclusions here, but the numbers fit into the transformation scenario.

\section{Tully-Fisher residuals}
\begin{figure}[]
   \centering
   \includegraphics[angle=0,width=\columnwidth]{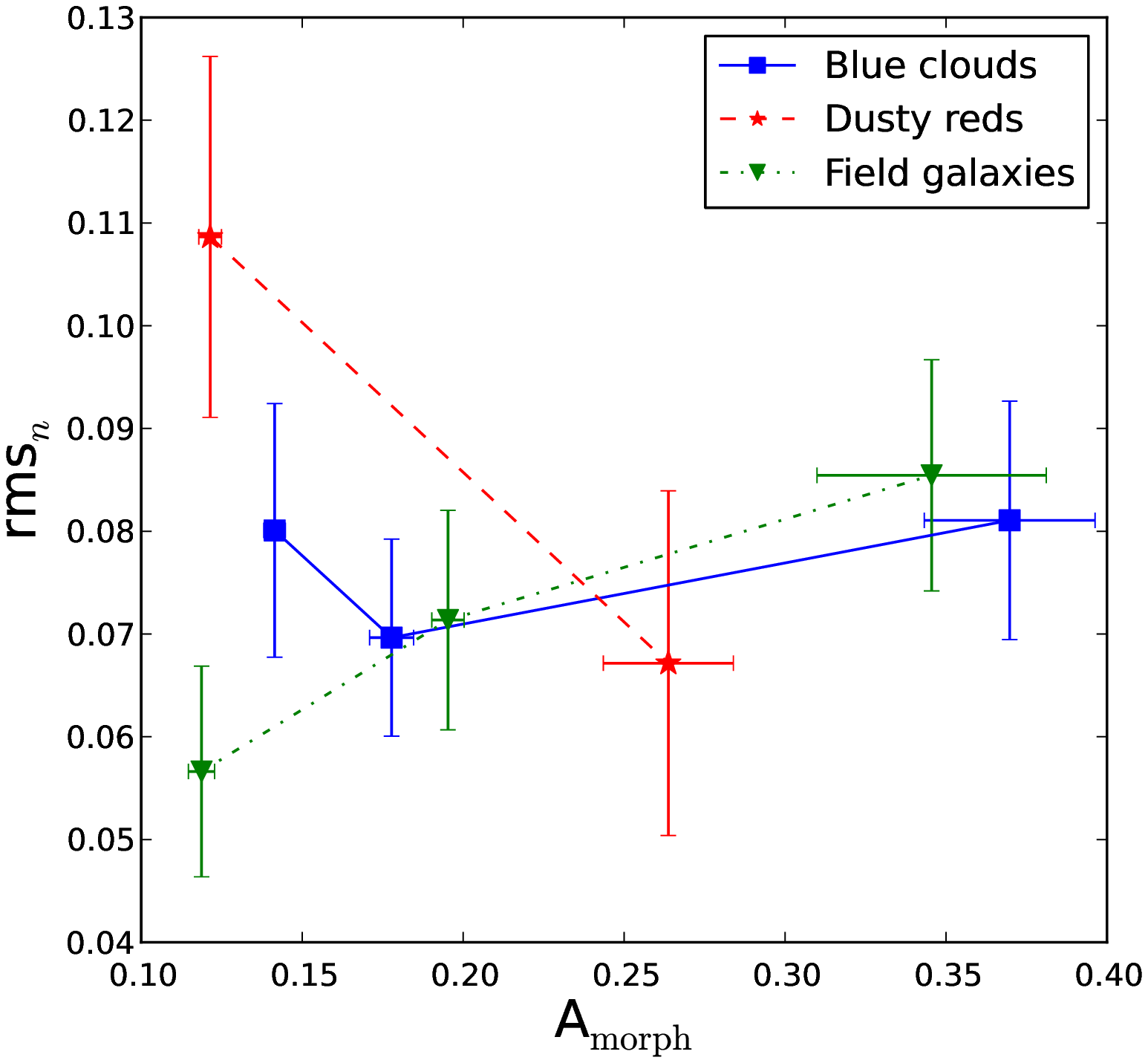}
      \caption{The $rms_n$ from RC fitting as a function of morphological asymmetry $A_{\mathrm{morph}}$ for blue cloud (43), dusty red (12) and field galaxies (57). Only high-quality objects are considered.
 }
         \label{fig:iA_rmsn}
   \end{figure}

\begin{figure*}[]
   \centering
   \includegraphics[angle=0,width=\textwidth]{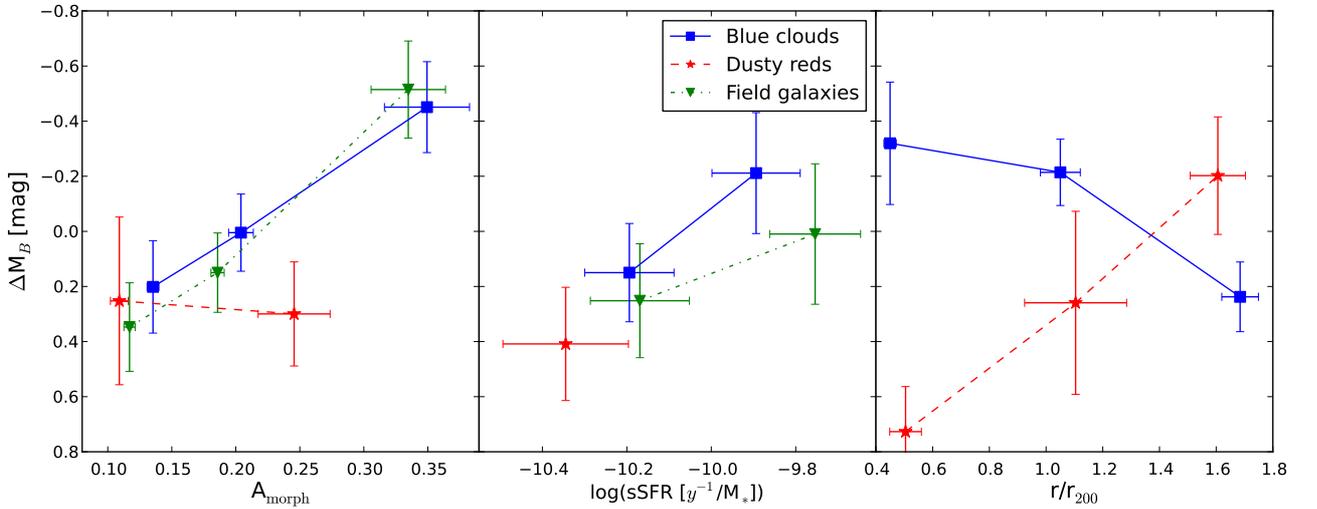}
      \caption{Mean residuals of the B-band TFR for the sample of high-quality cluster galaxies, separated into dusty red (red stars) and blue cloud galaxies (blue squares) as well as for the field sample (green triangles). Left panel: The correlation with morphological asymmetry is highly significant for blue cloud and field galaxies. Mid panel: Galaxies with high specific star-formation rates tend to lie above the TFR. Right panel: Dusty red ($\rho=-0.50$, $p=4.0\%$) and blue cloud ($\rho=0.39$, $p=0.5\%$) galaxies show opposite correlations as a function of cluster-centric distance.
 }
         \label{fig:resB}
   \end{figure*}

\begin{figure}[]
   \centering
   \includegraphics[angle=0,width=\columnwidth]{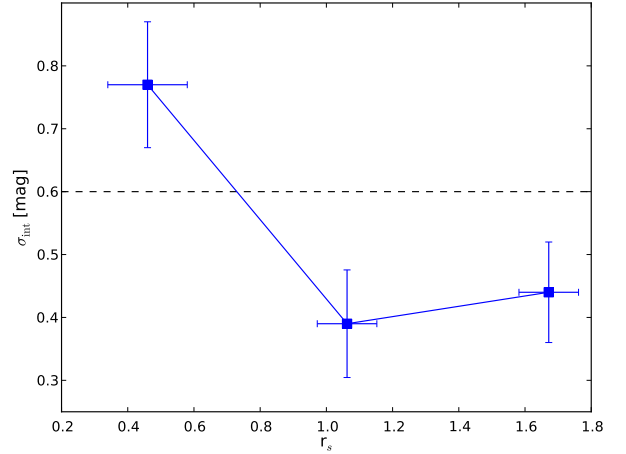}
      \caption{Intrinsic B-band scatter $\sigma_{\mathrm{int}}$ for three $r_s$-bins for the sample of high-quality cluster galaxies. The black dashed horizontal line depicts the intrinsic scatter of the total cluster sample.
 }
         \label{fig:RBrsabs}
   \end{figure}

\begin{figure}[]
   \centering
   \includegraphics[angle=0,width=\columnwidth]{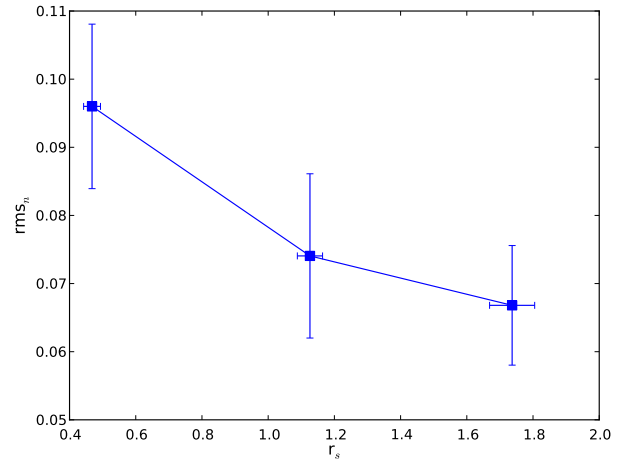}
      \caption{The $rms_n$ from RC fitting as a function of cluster-centric radius $r_s$ for the sample of high-quality cluster galaxies. The correlation is significant ($\rho=-0.30$, $p=4.3\%$).
 }
         \label{fig:rms_rs}
   \end{figure}

Several effects might influence the position of a galaxy in the TF plane. On the one hand, enhanced star-formation or higher dust-extinction rates can shift a galaxy along the luminosity axis, but do not affect stellar mass (changes in M/L and L cancel out). On the other hand, a high fraction of non-circular motion, possibly caused by kinematic distortions, can change the radial velocity profile and the measured rotational velocity. Variations in the stellar mass-to-light ratio or stellar-to-total mass ratio induce shifts in the TF plane \citep{kannappan02}.\\
In this section we mainly focus on the analysis of the B-band $\Delta M_{B}$ and stellar-mass residuals $\Delta M_{\ast}$. We examine their correlation with other parameters and a possible dependence on environment. We focus on the population of dusty red galaxies and investigate whether differences to blue cloud galaxies exist. As a proxy for environment we adopt the scaled cluster-centric radius $r_s = r/r_{\mathrm{200}}$ from Paper I. For a given subcluster, $r$ is the projected cluster-centric distance and $r_{\mathrm{200}}$ is the radius where the mean interior density is 200 times the critical density.\\
Furthermore, we also wish to have a measure for the asymmetry of the stellar disk at hand and we adopt the index $A_{\mathrm{morph}}$, which is computed by comparing an image to itself rotated by $180^{\circ}$ \citep{conselice00}. $A_{\mathrm{morph}}$ quantifies the degree of morphological distortions (see also Paper I). Fig. \ref{fig:iA_rmsn} shows the normalised rms as a function of the morphological asymmetry $A_{\mathrm{morph}}$. This plot resembles a plot from Paper I (Fig. 13 + 17), but this time the rotational asymmetry is parametrised by $rms_n$, i.e. the deviation of an observed rotation curve from its best-fit Courteau model. More kinematically distorted field galaxies exhibit higher morphological asymmetry. A Spearman rank-order correlation test yields a significant ($p=2.7\%$) correlation of $\rho=0.31$. On the other hand, dusty red galaxies ($\rho=-0.58$, $p=4.4\%$) have higher $rms_n$ for a low $A_{\mathrm{morph}}$. Although blue cloud galaxies show no significant correlation, they have - compared to field galaxies - also increased $rms_n$ values for smooth stellar disks. Following our argument in Paper I, interaction processes that affect the morphology (galaxy-galaxy interactions, harassment) most likely also distort the gaseous disk of a galaxy and hence perturb its rotation curve. Depending on the ICM density and the relative velocities, ram-pressure \citep[$\propto \rho_\mathrm{ICM}*v_{\mathrm{rel}}^{2}$,][]{gunn72} is able to significantly influence the gas content of a galaxy and strip part of its material, while not affecting its stellar disk \citep{kronberger08,chung08,kenney04,quilis00}. In the field, the $rms_n$ of the RC-fitting positively correlates with morphological disturbances. There, low relative velocities of galaxies favour tidal interactions. For morphologically undistorted cluster galaxies and especially dusty red galaxies a cluster-specific mechanism that effectively distorts the rotation curve but leaves the stellar disk unaffected is required. Ram-pressure stripping is the best candidate process.\\
Interestingly, low-quality and high-quality objects have the same mean morphological asymmetry ($A_{\mathrm{morph,lq}}=0.20\pm0.01$; $A_{\mathrm{morph,hq}}=0.22\pm0.02$). This is in agreement with \citet{jaffe11}, who also do not find a direct link between the morphological asymmetries in the stellar disk and the fraction of good rotation-curve fits.\\
This raises the question of whether the stellar asymmetry of a galaxy influences its offset from the best-fit TFR at all. Fig. \ref{fig:resB} (left panel) shows that for the class of high-quality objects a higher morphological asymmetry induces a brighter B-band residual. The correlations are highly significant ($p<0.1\%$) for the field ($\rho=-0.47$) and the cluster blue cloud ($\rho=-0.45$) population. Dusty red galaxies, on the other hand, produce no significant correlation ($\rho=0.20$, $p=27.9\%$). The same behaviour is seen for low-quality objects (not shown in the figure).\\
 Galaxies with a higher morphological asymmetry index tend to have a more late-type SED and higher star-formation rates, which, in turn, leads to brighter residuals. Also \citet[e.g.][]{giovanelli97} found TF offsets slightly fainter for more early-type galaxies. This could explain part of these correlations. On the other hand, galaxy-galaxy interactions, accompanied by morphological distortions, can induce a star-formation enhancement and consequently result in larger TF offsets \citep[e.g.][]{alonso04,lambas03}.\\
Fig. \ref{fig:resB} (mid panel) confirms that galaxies with higher specific star-formation rates preferentially populate the region above the B-band TFR. Dusty red galaxies have reduced specific star formation rates. Note that we only have robust star-formation rates (determined from UV and infra-red data) of 30/55 high-quality cluster (22/43 blue clouds; 8/12 dusty reds) and 22/57 high-quality field galaxies. Our finding is in agreement with \citet{kannappan02} who also detect a correlation between \ion{H}{$\alpha$} equivalent widths (a proxy for specific star formation) and TF residuals.\\
We discuss possible environmental effects on the B-band TF residuals in Fig. \ref{fig:resB} (right panel). The mean B-band residuals of dusty red ($\rho=-0.50$, $p=4.0\%$) and blue cloud ($\rho=0.39$, $p=0.5\%$) galaxies show a different behaviour as a function of cluster-centric radius. While dusty red galaxies become fainter towards the cluster center, blue cloud galaxies tend to be more luminous at small $r_s$. An enhanced ICM density and ram pressure may trigger star formation in the gas-rich blue cloud population \citep{kronberger08b,steinhauser12}, while the gas-poor dusty red galaxies might already be depleted of most of their gas due to ram-pressure stripping.\\
The denser environment seems to affect the residuals of cluster galaxies. Galaxies closer to the cluster centre deviate more from the TFR, which is confirmed by the fact that the absolute B-band residuals of high-quality objects increase towards the cluster centre ($\rho=-0.27$, $p=1.9\%$). Basically, higher absolute residuals produce a larger scatter. Fig. \ref{fig:RBrsabs} shows the intrinsic B-band scatter of high-quality objects for three cluster-centric radius bins. Since we detect the same trend for dusty red and blue cloud galaxies we only show the combined cluster sample. The TFR of galaxies at lower $r_s$ has a larger intrinsic scatter. This might be due to a mix of several effects. Gas-rich (blue cloud) galaxies have increased star-formation rates towards the cluster centre, while gas-poor (dusty red) galaxies are located below the TFR, possibly due to quenching in progress. The observed trends in the B-band residuals are not significant for the corresponding stellar-mass residuals. Hence, these trends in the B-band probably are mainly due to M/L-variations. A higher amount of RC distortions may also be responsible for the larger scatter at smaller $r_s$: galaxies residing closer to the cluster center have on average higher $rms_n$ values ($\rho=-0.30$, $p=4.3\%$, see Fig. \ref{fig:rms_rs}), i.e. their rotation curves deviate more form a Courteau model.\\

\section{Summary and conclusions}
We continue the work of Paper I based on long-slit spectroscopy of disk galaxies around the multiple cluster system Abell 901/902 taken with the VLT instrument VIMOS. We exploit up to four emission lines per object to extract rotation curves of 96 cluster and 86 field galaxies. Assuming a tilted ring model and using input parameters such as the inclination angle, the location of the apparent major axis with respect to the direction of dispersion, the width and tilt angle of the slit, and the FWHM of the seeing disk, we simulate long-slit spectroscopy. Adopting a Courteau model for the radial rotation law of a galaxy, this simulation enables us to find the best-fit intrinsic maximum rotation velocity given the observed rotation curve and the input parameters. We assess the quality of a rotation curve visually (distinguishing between high-quality and low-quality objects) and by utilising the rms of the rotation-curve fitting. We then use the rotation velocity to build a B-band Tully-Fisher relation (with absolute magnitudes derived from HST/ACS V-band images) and a stellar-mass Tully-Fisher relation. We fit a regression line adopting a maximum likelihood method based on Bayes' theorem. Eventually, we analyse the TF relations and examine B-band and stellar-mass residuals. We focus on differences between cluster and field environment, or between dusty red and blue cloud galaxies in the cluster. Dusty red galaxies are mainly found in the cluster environment and populate the red sequence in an U-V colour-magnitude diagram. This is partly due to significant dust extinction and partly due to reduced star formation rates \citep{wolf05}. The main results of this work are:  
\begin{itemize}
      \item[1.]
 The best-fit TF parameters (slope, intercept, intrinsic scatter) between cluster and field galaxies agree within the error bars. However, if we assume a fixed slope, high-quality field galaxies are $\Delta M_{B}=-0\fm42\pm0\fm15$ brighter than the corresponding cluster population. Accounting for the higher redshift and on average bluer SED of the field population explains half of this effect. Since we find no offset for the stellar-mass TFR, the residual difference might stem from M/L-variations due to environmental effects.
  \item[2.]
 When considering low-quality objects only, the intrinsic scatter is increased for cluster galaxies compared to field galaxies. This indicates more frequent kinematic distortions in the cluster environment.
  \item[3.]
  Compared to local TFRs we obtain a higher intrinsic scatter and a slightly shallower slope. At fixed slope, we find a B-band offset of $\Delta M_{B}\lesssim-0\fm33\pm0\fm10$. Since we find no difference in the stellar-mass intercept, this is, at least in part, due to a luminosity evolution with look-back time ($\sim 2$ Gyr).
   \item[4.]
   While there is no significant difference in the average stellar-mass residuals, dusty red galaxies are $\Delta M_{B}=0\fm35\pm0\fm12$ fainter than blue cloud galaxies for a given velocity. This indicates different gas contents between the two SED-types.
  \item[5.]
  Exploiting the normalised rms of the RC fitting, i.e. the deviation of a RC from its best-fit Courteau model, field galaxies are kinematically more distorted towards higher morphological asymmetry. In contrast, cluster galaxies (and especially dusty red galaxies) with a smooth stellar disk show increased kinematic disturbances compared to field galaxies. A cluster-specific interaction mechanism like ram-pressure stripping, which does not affect the morphology but only the kinematics, is the most likely explanation for this trend.
    \item[6.]
    Galaxies with higher specific star-formation rates have on average higher B-band residuals.
     \item[7.]
     Dusty red galaxies are fainter towards the cluster centre, whereas the B-band residuals of blue cloud galaxies become brighter for lower cluster-centric radii. While the SFR of gas-rich galaxies might be enhanced in denser environments, the gas-poor dusty red galaxies might be in the progress of quenching due to ram-pressure stripping when falling into the cluster.
	\item[8.]
	Galaxies residing close to the cluster centre show a larger TFR scatter. This is probably caused by increased/decreased star-formation rates and/or distorted kinematics.
\end{itemize}
Altogether, the results of the TF analysis are consistent with the scenario presented in Paper I, which suggests that cluster galaxies are affected by the dense environment they are residing in. Ram-pressure is an important mechanism especially for dusty red galaxies, which might be an intermediate stage in the transformation of infalling field spiral into cluster lenticular galaxies.

\begin{acknowledgements}
      We thank ESO for the support during the spectroscopic observations. We thank the anonymous referee
for improving the quality of this work. For plotting, the 2D graphics environment \textit{Matplotlib} is used \citep{hunter07}. We calculate cosmological values such as distance modulus and angular scale using the prescriptions of \citet{wright06}. Asmus B\"{o}hm and Sabine Schindler thank the Austrian Science Fund (FWF) for funding (projects P19300-N16, P23946-N16).
\end{acknowledgements}

\bibliographystyle{aa}
\bibliography{benni}

\end{document}